\documentclass[tikz]{amsart}
\usepackage[dvips]{graphicx}
\usepackage{amsmath,amssymb,amsthm,bm,fancybox,fullpage,dashbox}
\usepackage{bm}
\usepackage[all]{xy}
\usepackage{cite}
\usepackage{verbatim}
\usepackage{enumerate}
\usepackage{url}
\usepackage{color}
\usepackage{tikz-cd, mathdots}

\usepackage{setspace}

\theoremstyle{definition}

\newtheorem{theorem}{Theorem}[section]

\newtheorem{definition}[theorem]{Definition}
\newtheorem{remark}[theorem]{Remark}
\newtheorem{example}[theorem]{Example}

\theoremstyle{remark}

\newcommand{\N}{\mathbb{N}}

\newcommand{\ra}{\rightarrow}

\newcommand{\xra}[1]{\xrightarrow{#1}}
\newcommand{\lra}{\longrightarrow}

\newcommand{\aut}{\mathsf{Aut}}
\newcommand{\stext}[1]{\overline{#1}}
\newcommand{\PSS}{\mathsf{PSS}}
\newcommand{\turn}{\mathsf{turn}}
\newcommand{\shuffle}{\mathsf{shuffle}}
\newcommand{\F}{\mathcal{F}}

\newcommand{\act}{\mathsf{act}}

\newlength{\ralen}
\newcommand{\back}{\raisebox{\ralen}{\framebox(11,13){{\large \textbf{?}}}}}

\newcommand{\crd}[1]{\raisebox{\ralen}{\framebox(11,13){#1}}}

\newcommand{\md}[1]{\textcolor{black}{#1}}
\newcommand{\f}[1]{\mathsf{#1}}
\definecolor{Green}{rgb}{0.1, 0.5, 0.1}

\begin{document}
\title{Graph Automorphism Shuffles from Pile-Scramble Shuffles}

\author[K. Miyamoto]{Kengo Miyamoto}
\author[K. Shinagawa]{Kazumasa Shinagawa}
\address[K. Miyamoto]{Ibaraki University, 4-12-1 Nakanarusawa, Hitachi, Ibaraki, 316-8511, Japan.}
\email{kengo.miyamoto.uz63@vc.ibaraki.ac.jp}
\address[K. Shinagawa]{Ibaraki University, 4-12-1 Nakanarusawa, Hitachi, Ibaraki, 316-8511, Japan; National Institute of Advanced Industrial Science and Technology (AIST), Tokyo Waterfront Bio-IT Research Building 2-4-7 Aomi, Koto-ku, Tokyo, 135-0064, Japan.}
\email{shinagawakazumasa@gmail.com}

\maketitle
\begin{abstract}
A pile-scramble shuffle is one of the most effective shuffles in card-based cryptography. 
Indeed, many card-based protocols are constructed from pile-scramble shuffles. 
This article aims to study the power of pile-scramble shuffles. 
In particular, for any directed graph $G$, we introduce a new protocol called ``a graph shuffle protocol for $G$'', and show that it \md{can be implemented} by using pile-scramble shuffles only. 
Our proposed protocol requires $2(n+m)$ cards, where $n$ and $m$ are the numbers of vertices and arrows of $G$, respectively. 
The number of pile-scramble shuffles is $k+1$, where $1 \leq k \leq n$ is the number of distinct degrees of vertices of $G$. 
As an application, a random cut for $n$ cards, which is also an important shuffle, can be realized by \md{$3n$} cards and \md{two} pile-scramble shuffles. 
\end{abstract}

\section{Introduction}

\subsection{Background}

Let $X, Y$ be finite sets, $n \in \N$ a natural number, and $f: X^n \ra Y$ a function. 
Suppose that $n$ players each having $x_i \in X$ as input wish to know an output value $f(x_1, x_2, \ldots, x_n) \in Y$ without \md{revealing anything about their own inputs beyond the output value to other players}. 
Secure computation protocols can solve this kind of situation. 
Secure computation, which was formalized by Yao\cite{Yao82, Yao86}, plays an important role in cryptography (cf. see the survey on secure computation by Lindell \cite{Lindell20}). 

Card-based cryptography \cite{Boer89, Kilian93} is a kind of secure computation, which uses a deck of physical cards. 
Given a sequence of face-down cards (which is typically an encoding of input $(x_1, x_2, \ldots, x_n) \in X^n$), a card-based protocol transforms it to an output sequence (which is typically an encoding of output $f(x_1, x_2, \ldots, x_n) \in Y$) by a bunch of physical operations on cards. 
One of the features of card-based cryptography is that it allows us to understand intuitively the correctness and security of a protocol, since we can actually perform the protocol by hands. 
For this reason, it is expected to be used as an educational material. 
Indeed, some universities \cite{CHL13, Marcedone15, MizukiTeaching16} have actually \md{used} card-based cryptography as an educational material. 

In card-based protocols, a \emph{shuffle}, which is a probabilistic rearrangement, is allowed to \md{apply to a sequence of cards}. 
It is considered as the most crucial operation in card-based protocols since randomness from shuffles is the primary tool to obtain the security of protocols. 
Among shuffles, a \emph{(pile) random cut (RC)}, a \emph{random bisection cut (RBC)}, and a \emph{pile-scramble shuffle (PSS)} are the most effective shuffles\footnote{\md{Our classification focuses on the group structure of permutations: the cyclic groups (RCs) and the symmetric groups (PSSs). Since RBCs are historically important shuffles and the intersection of RCs and PSSs, we classify them as RC, RBC, and PSS.}} \md{in card-based cryptography}. 
Indeed, most card-based protocols are constructed with these shuffles only (cf. protocols with RCs only \cite{Boer89, Kilian93, Niemi98, Niemi99, Stiglic01, Uchiike06, Heather14, MiyaharaCOCOA18, ShinagawaICISC18, ToyodaAPKC20, SaitoTPNC20, MiyaharaTCS20, KoyamaAPKC21, RobertCiE21, AbeNGC21, KochFUN21}, protocols with RBCs only \cite{MizukiFAW09, MizukiAC12, MizukiUCNC13, NishidaTPNC13, NishidaTAMC15, NishidaIEICE15, MizukiCANS16, ShinagawaFAW19,RuangwisesTCS21}, protocols with PSSs only \cite{GradwohlFUN07, PascalFUN16, SasakiFUN18, Koch18b, RobertSSS20, SasakiTCS20, ShinagawaDAM21, MurataWALCOM21,RuangwisesNGC21}, protocols with RCs and RBCs only \cite{AbeAPKC18, TakashimaTPNC19, KoyamaCSR21, MiyaharaFUN21}, protocols with RCs and PSSs only \cite{ShinagawaSSS18, TakashimaCOCOA19, DumasCOCOON19, MiyaharaISPEC19, TakashimaTCS20, RuangwisesTCS21a}, and protocols with RBCs and PSSs only \cite{IshikawaUCNC15, MizukiTCS16, HashimotoICITS17, ShinodaSecITC20}). 
With this background, it is essential to study further what can be done by these shuffles. 

\subsection{Contribution}

In this \md{paper}, we show that \emph{graph shuffles} can be \md{implemented} with PSSs. 
Let $G$ be a directed graph\footnote{\md{We regard undirected graphs as directed graphs by identifying each undirected edge with two directed edges with opposite directions.}}. 
A graph shuffle for $G$ is a shuffle that arranges a sequence of cards according to an automorphism of $G$ chosen uniformly at random. 
Our main contribution is to construct a card-based protocol that achieves a graph shuffle for any graph $G$. 
We call this a \emph{graph shuffle protocol for $G$}. 
The number of cards in our protocol is $2(n+m)$, where $n$ and $m$ are the numbers of vertices and \md{edges} of $G$, respectively. 
\md{The number of shuffles (i.e., PSSs) in our protocol is $|\f{Deg}_G|+1$, where $\f{Deg}_G$ is the set of vertex degree of $G$ (see Section \ref{ss:graph}).}
\md{We remark that our protocol has one drawback: it requires to compute a graph isomorphism between $G$ and its isomorphic graph $G'$. In general, computing a graph isomorphism is a complex computational task (see also Remark \ref{rem:iso}). We conjecture that computing a graph isomorphism is inherent in implementing a graph shuffle. We left it as an open problem whether computing a graph isomorphism can be removed or not.}

A class of graph shuffles includes many interesting shuffles (see Section \ref{ss:implication}). 
Indeed, RCs, RBCs, and PSSs are special cases of graph shuffles. 
\md{In particular, a RC is a graph shuffle for a directed cycle graph.} 
A straightforward corollary of our main result is that a RC can be implemented with PSSs. 
Since a PSS can be implemented with RCs (cf. see Cr\'{e}peau and Kilian \cite{Kilian93}'s idea for generating a random fixed-free permutation), PSSs and RCs are essentially equivalent from the viewpoint of feasibility. 
\md{It is worthwhile to mention the importance of the fact that RCs are implementable by PSSs. From the theoretical viewpoint, this shows that every protocol with RCs is transformed into a protocol with PSSs and vice versa. 
From the practical viewpoint, you can choose whether to use RCs or PSSs as shuffles in a protocol execution. 
In order to execute a RC by hand, we need to ensure that everyone must be able to verify that the rearrangement is indeed a cyclic shift while hiding the rearrangement itself. 
On the other hand, a PSS can be done by a rearrangement of piles in a completely randomly fashion although it requires physical envelopes as an additional tool. 
Which shuffle can be easily executable depends on a situation and thus there should be some cases that PSSs are more desirable than RCs. 
}

\md{Due to the importance of the result of RC, we improve a graph shuffle protocol for a directed cycle graph. 
In particular,} for the directed cycle graph with $n$ vertices, we design a graph shuffle protocol with \md{$3n$} cards while the general protocol requires \md{$4n$} cards. 

\md{We also improve a graph shuffle protocol for an undirected cycle graph. 
A graph shuffle for the undirected cycle graph is equivalent to the \emph{dihedral shuffle}, which is introduced by Niemi and Renvall \cite{Niemi98}.} For the undirected cycle graph with $n$ vertices, we design a graph shuffle protocol with \md{$3n$} cards while the general protocol requires \md{$6n$} cards. 

\subsection{Related works}\label{relatedworks}

\md{
Koch and Walzer \cite{KochFUN21} showed that \emph{uniform closed shuffles} (see Definition \ref{def:ucshuffle}) can be implemented with RCs only. 
It is an essential milestone for implementing uniform closed shuffles. 
Since graph shuffles are uniformly closed, Koch and Walzer's method allows that every graph shuffle can be done by RCs.
However, we point out that their protocol requires each party somehow to generate a uniformly random element of a given group in the party's head. This action is not allowed in the Mizuki-Shizuya model \cite{MizukiIJISEC14} which is known as the standard computational model of card-based cryptography. 
From this viewpoint, our protocol for graph shuffles and their protocol are based on different models of card-based cryptography. 
Our motivation is to implement a subclass of uniform closed shuffles in the Mizuki-Shizuya model. 
Besides the theoretical aspect, it is worthwhile to note that removing a randomness generation in the head brings a practical benefit for security because it is not clear how close the distribution of random elements generated in the head will be to the distribution of truly random elements. 
}




\section{Preliminaries}

In this section, we collect some fundamentals in card-based cryptography; see \cite{MizukiIJISEC14} for example.

\subsection{Cards}\label{ss:card}
Throughout \md{this paper}, we deal with physical \textit{cards} with the symbol ``?"  on the backs. 
\md{We use two collections of cards: \textit{black-cards} $\crd{1}\,\crd{2}\,\crd{3}\,\cdots$ and \textit{red-cards} $\crd{$\stext{1}$}\,\crd{$\stext{2}$}\,\crd{$\stext{3}$}\,\cdots$ as follows: }
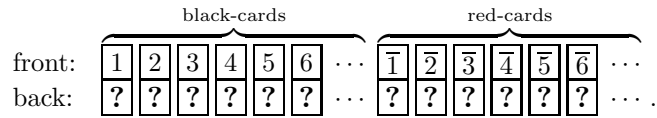
\[ \begin{array}{ll}
\text{front:} & \overbrace{\crd{1}~\crd{2}~\crd{3}~\crd{4}~\crd{5}~\crd{6}~\cdots}^{\text{black-cards}}~\overbrace{\crd{$\stext{1}$}~\crd{$\stext{2}$}~\crd{$\stext{3}$}~\crd{$\stext{4}$}~\crd{$\stext{5}$}~\crd{$\stext{6}$}~\cdots}^{\text{red-cards}}\, \\
\text{back:} & \back~\back~\back~\back~\back~\back~\cdots~\back~\back~\back~\back~\back~\back~\cdots.
\end{array}
\]
We distinguish between the natural number $\stext{i}$ (written in red) and the natural number $i$ (written in black). 
We denote by $\mathbb{N}^{\f{red}}$ the set of all natural numbers written in red, i.e., $\mathbb{N}^{\f{red}} = \{\stext{1}, \stext{2}, \stext{3}, \ldots\}$.
The set $\mathbb{N}^{\f{red}}$ is a totally ordered set by using the natural order on $\mathbb{N}$.
We define a totally order $\preccurlyeq$ on $\mathbb{N}\cup\mathbb{N}^{\f{red}}$ by
$\alpha \preccurlyeq \beta$ if and only if 
\begin{itemize}
\item $x, y \in \mathbb{N}$ and $x\leq y$, where $\alpha = x$ and $\beta = y$, 
\item $\stext{x}, \stext{y}\in\mathbb{N}^{\f{red}}$ and $x\leq y$, where $\alpha = \stext{x}$ and $\beta = \stext{y}$, or
\item $\stext{x}\in \mathbb{N}^{\f{red}}$ and $y\in\mathbb{N}$, where $\alpha = \stext{x}$ and $\beta = y$.
\end{itemize}
\md{\textit{A deck} $D$} is a non-empty multiset such that $\{?\}\cap D=\varnothing$. 
Let $D$ be a deck.  
An expression $\dfrac{x}{?}$ $\left(\text{resp. } \dfrac{?}{x}\right)$ with $x\in D$ is said to be \textit{a face-up card} (resp.  \textit{a face-down card}) of $D$.
\textit{A lying card} $y$ of $D$ is the face-up card $y=\dfrac{x}{?}$ of $D$ or the face-down card $y=\dfrac{?}{x}$ of $D$, and in this case, we set $\f{atom}(y)=x$.
\textit{A card-sequence} from $D$ is a list of lying cards of $D$, say $(x_1,\ldots,x_n)$, such that $\{\f{atom}(x_i)\mid i=1,2,\ldots,n\}=D$ as multisets.
For a card-sequence $\f{x}$, we write $\f{x}_i$ for the $i$-th term. 
A face-up card $\dfrac{x}{?}$ is represented by \crd{$x$}~, and a face-down card $\dfrac{?}{x}$ is represented by $\back$~.
Given a card $x$ with the expression $\dfrac{y}{z}$, we write $\f{front}(x)=y$, $\f{back}(x)=z$, and $\f{swap}(x)=\dfrac{z}{y}$.
For a card-sequence $\f{x}=(\f{x}_1,\ldots,\f{x}_n)$ and a subset $T\md{\subseteq} \{1,2,\ldots, n\}$, we define an operator $\turn_T(-)$ by
\[ 
\f{turn}_T(\f{x})=(y_1,\ldots, y_n), \quad y_i=
\left\{\begin{array}{ll}
\f{swap}(\f{x}_i) & \text{if $i\in T$,}\\
\f{x}_i  &\text{if $i\notin T$.}
\end{array}\right.
\]
The card-sequence $\f{front}(\f{x}) = (\f{front}(\f{x}_1),\ldots, \f{front}(\f{x}_n))$  is called \textit{the visible sequence} of $\f{x}$.
Let $(\mathcal{T}, \mathcal{G})$ be a pair of a collection of subsets of $\{1,2,\ldots, n\}$ (i.e., $\mathcal{T} \md{\subseteq} 2^{\{1,2,\ldots, n\}}$) and a probability distribution on $\mathcal{T}$.
Now, we also define an operation $\f{rflip}_{(\mathcal{T},\mathcal{G})}(-)$ associated with the pair $(\mathcal{T}, \mathcal{G})$ by
\[ \f{rflip}_{(\mathcal{T},\mathcal{G})}(\f{x})= \f{turn}_T(\f{x}), \]
where $T$ is chosen from $\mathcal{T}$ depending on the probability distribution $\mathcal{G}$.
\md{Note that if $\mathcal{T}=\{T\}$ with a subset $T\md{\subseteq} \{1,2,\ldots, n\}$,} then $ \f{rflip}_{(\mathcal{T},\mathcal{G})}(-)=\f{turn}_T(-)$.

\subsection{Shuffles}\label{ss:shuffle}
For a natural number $n\in\mathbb{N}$, we denote by  $\mathfrak{S}_n$ the symmetric group of degree $n$, that is, the group whose elements are all bijective maps from $\{1,2,\ldots, n\}$ to itself, and whose group multiplication is the composition of functions. An element of the symmetric group is called \textit{a permutation}.

Given a card-sequence  $\mathsf{x}=(\f{x}_1,\ldots,\f{x}_n)$ and $\sigma\in\mathfrak{S}_n$, we have a card-sequence $\sigma(\f{x})$ in the natural way:
\[ \sigma(\mathsf{x})=(\f{x}_{\sigma^{-1}(1)},\ldots, \f{x}_{\sigma^{-1}(n)}).\]

Now, we recall an operation on a card-sequence which is called a ``shuffle".
Roughly speaking, a shuffle is a probabilistic reordering operation on a card-sequence. 
Let $(\Pi,\F)$ be a pair of a subset of $\mathfrak{S}_n$ and a probability distribution on $\Pi$. 
For a card-sequence $\f{x}=(\f{x}_1,\ldots,\f{x}_n)$, an operation $\shuffle_{(\Pi,\F)}(-)$ associated with the pair $(\Pi,\F)$ is defined by
\[ \shuffle_{(\Pi,\F)}(\f{x})= \sigma(\f{x}).  \]
\md{Here, $\sigma$ is chosen according to the probability distribution $\mathcal{F}$ on $\Pi$.}
Note that when we apply a shuffle to a card-sequence, no one knows which permutation was actually chosen.
We also note that if $\Pi=\{\sigma\}$ for some $\sigma\in\mathfrak{S}_{n}$, then $ \f{shuffle}_{(\Pi,\mathcal{F})}(-)=\sigma(-)$. 

\begin{definition}\label{def:ucshuffle}
A shuffle $\shuffle_{(\Pi,\F)}$ is said to be \textit{uniform closed} if $\Pi$ is closed under the multiplication of the symmetric group, and $\F$ is the uniform distribution on $\Pi$. 
\end{definition}

All shuffles dealt with \md{this paper} are uniform closed shuffles. 

\begin{example}
\begin{enumerate}
\item[(1)] \md{
For a sequence of $\ell$ cards, suppose that a subsequence of the sequence is divided into $n$ piles of $m$ cards. (It holds $\ell \geq nm$.) 
\textit{A pile-scramble shuffle} (PSS for short) is a uniform closed shuffle that completely randomly permutes $n$ piles.}
The following shuffle is an example of a PSS:
\[\mathsf{PSS}_{(3,2)}:
\left(~\crd{1}~\crd{2}~,~\crd{3}~\crd{4}~,~\crd{5}~\crd{6}~ \right) \overset{\sigma}{\longmapsto}
\begin{cases}
\left(~\crd{1}~\crd{2}~,~\crd{3}~\crd{4}~,~\crd{5}~\crd{6}~ \right)& \text{if $\sigma=\mathsf{id}$,}\\
\left(~\crd{1}~\crd{2}~,~\crd{5}~\crd{6}~,~\crd{3}~\crd{4}~ \right)& \text{if $\sigma=\mathsf{(2~3)}$,}\\
\left(~\crd{3}~\crd{4}~,~\crd{5}~\crd{6}~,~\crd{1}~\crd{2}~ \right)&\text{if $\sigma=\mathsf{(1~3~2)}$,}\\
\left(~\crd{3}~\crd{4}~,~\crd{1}~\crd{2}~,~\crd{5}~\crd{6}~ \right)& \text{if $\sigma=\mathsf{(1~2)}$,}\\
\left(~\crd{5}~\crd{6}~,~\crd{1}~\crd{2}~,~\crd{3}~\crd{4}~ \right) &\text{if $\sigma=\mathsf{(1~2~3)}$,}\\
\left(~\crd{5}~\crd{6}~,~\crd{3}~\crd{4}~,~\crd{1}~\crd{2}~ \right) & \text{if $\sigma=\mathsf{(1~3)}$.}\\
\end{cases}
\]
We use $\mathsf{PSS}_{(n,m)}$ to denote a PSS for $n$ piles each having $m$ cards. 
\md{We remark that $\mathsf{PSS}_{(n,m)}$ can be easily implemented by putting each pile into each physical envelope
 and then permute them.}

\item[(2)] Let $\pi_k\in\mathfrak{S}_n$ be the permutation
\[\pi_k=\begin{pmatrix}
1 & 2 & \cdots &  k & k+1 & \cdots & n \\
n-k+1 & n-k+2 & \cdots & n & 1 & \cdots & n-k
\end{pmatrix},\]
and set $\Pi=\{\pi_k\mid k=1,2,\ldots,n\}$. \md{This} uniform closed shuffle $\shuffle_{(\Pi,\F)}$ is called \textit{a \md{random} cut} (RC for short). 
\end{enumerate}
\end{example}

\subsection{Procotols}\label{ss:protocol}
Mizuki and Shizuya \cite{MizukiIJISEC14} define the formal definition of a card-based protocol via an abstract machine.
In this section, we recall the definition of a card-based protocol and introduce a shuffle protocol, which is a particular card-based protocol realizing a shuffle. 

\subsubsection{Card-based protocols}\label{sss:reduction}

To put it briefly, a ``protocol" is a Turing machine that chooses one of the following operations to be applied to a card-sequence $\f{x}$: turning $(\f{x}\mapsto \f{rflip}_{(\mathcal{T},\mathcal{G})}(\f{x}))$ or shuffling $(\f{x}\mapsto \shuffle_{(\Pi,\F)}(\f{x}))$.

For a deck $D$, the set of all card-sequences from $D$ will be denoted by $\f{Seq}^D$.
Then \textit{the visible sequence set} $\f{Vis}^D$ is defined as the set of all sequences $\f{front(x)}$ for $\f{x}\in \f{Seq}^D$.
We also define the sets of the actions:
\begin{align*}
& \f{turn}^n=\{\f{turn}_T(-)\mid T\md{\subseteq} \{1,2,\ldots, n\}\}, \\
& \f{perm}^n=\{\sigma(-)\mid \sigma\in\mathfrak{S}_n\}, \\
& \f{SP}^n=\{\shuffle_{(\Pi,\mathcal{F})}(-)\mid \text{$\mathcal{F}$ is  a probability distribution on $\Pi\in 2^{\mathfrak{S}_n}\}, \text{ and}$}\\
& \f{TP}^n=\{\f{rflip}_{(\mathcal{T},\mathcal{G})}(-)\mid \text{$\mathcal{G}$ is a probability distribution on $\mathcal{T}\md{\subseteq} 2^{\{1,2,\ldots, n\}}\}$}.
\end{align*}

A protocol is a Markov chain, that  is, a stochastic model describing a sequence of possible actions in which the probability of each action depends only on the state attained in the previous event.
Let $Q$ be a finite set with two distinguished states, which are called \textit{an initial state} $q_0$ and \textit{a final state} $q_{\rm f}$.

\begin{definition}
\textit{A card-based protocol} is a quadruple $\mathcal{P}=(D,U,Q,\f{A})$, where $U\md{\subseteq} \f{Seq}^D$ is an input set and $\f{A}$ is a partial action function
\[ 
\begin{array}{cccc}
 \f{A}:  & (Q\setminus\{q_{\rm f}\})\times \f{Vis}^D &  \longrightarrow &  Q\times (\f{turn}^n\cup\f{perm}^n\cup \f{SP}^n\cup \f{TP}^n), \\
           & (q,\f{y})                                                    & \longmapsto & (q',\f{act}_{q,\f{y}})
 \end{array} \]
which depends only on the current state and visible sequence, specifying the next state and an operation on the card-sequence from $(\f{turn}^n\cup\f{perm}^n\cup \f{SP}^n\cup \f{TP}^n)$, such that $\f{A}(q_0,\f{front(x)})$ is defined if $\f{x}\in U$.
For a state $q\in Q\setminus\{q_{\rm f}\}$ and a visible sequence $\f{y}=\f{front(x)}\in \f{Vis}^D$ such that $\f{A}(q,\f{y})=(q',\f{act}_{q,\f{y}})$, we obtain the next state $(q' ,\f{front}(\act_{q,\f{y}}(\f{x})))$.
\md{By the above process, if we have $(q_{\rm f},\mathsf{X}) \in Q\times\f{Vis}^D$ for some $\mathsf{X}\in \f{Vis}^D$,}
the protocol $\mathcal{P}$ terminates. 
\end{definition}

Let $\mathcal{P}=(D,U,Q,\f{A})$ be a card-based protocol.
For an execution of $\mathcal{P}$ with an input card-sequence $\f{x}^{(0)}\in U$, we obtain a sequence of results of actions as follows:
\[ (q_0,\f{x}^{(0)})\longmapsto (q_1,\f{x}^{(1)})\longmapsto (q_2,\f{x}^{(2)})\longmapsto(\md{q_3},\f{x}^{(3)})\longmapsto \cdots,\]
where $\f{x}^{(i)}=\act_{q_{i-1},\f{front}(\f{x}^{(i-1)})}(\md{\f{x}^{(i-1)}})$ for $i\geq 1$. \md{Here, $q_i$ ($i = 0, 1, 2, \ldots$) are not necessarily distinct}. 
If \md{the action function value} $\f{A}(q_i,\f{x}^{(i)})$ is undefined for some $i \in \N$, we say that ``$\mathcal{P}$ aborts at Step $i$ in the execution". 
Note that even for the same input card-sequence $\f{x}^{(0)}$, the obtained chains may be different for each execution.
If the protocol $\mathcal{P}$ terminates for an input card-sequence $\f{x}^{(0)}$, then we have a chain of results as follows:
\[ (q_0,\f{x}^{(0)})\longmapsto (q_1,\f{x}^{(1)})\longmapsto (q_2,\f{x}^{(2)})\longmapsto(q_2,\f{x}^{(3)})\longmapsto \cdots \longmapsto (q_{\rm f},\f{x}^{(\ell)}).\]
In this case, $\f{x}^{(0)}$ is called \textit{an initial sequence}, $\f{x}^{(\ell)}$ is called \textit{a final sequence}, and the sequence 
\[ (\f{y}^{(0)},\f{y}^{(1)},\ldots, \f{y}^{(\ell)}), \]
where $\f{y}^{(i)}=\f{front}(\f{x}^{(i)})$, is called \textit{a visible sequence-trace} of $\mathcal{P}$. 
We denote by $\f{Fin}(\mathcal{P})$ the set of  all \md{final} sequences, which is obtained by $\mathcal{P}$.


\begin{example}
Let us consider the following. Take the deck $D=\{1,2,3,4\}$, and hence use as follows:
\[ \begin{array}{ccccc}
\text{front:} & \crd{1}~& \crd{2}~ & \crd{3}~ & \crd{4}~\;\\
\text{back:} & \back~ & \back~   &  \back~  & \back~.
\end{array}
\]
Now, we give a card-based protocol $\mathcal{P}=\left(D, \left\{\left(\dfrac{1}{?},\dfrac{2}{?},\dfrac{3}{?},\dfrac{4}{?}\right)\right\}, \{q_0,q_1,q_2,q_{\rm f}\}, \f{A}\right)$ such that
\begin{align*}
&\f{A}\left(q_0,1234\right) = (q_1, \f{turn}_{\{1,2,3,4\}}(-)), \\
&\f{A}\left(q_1,????\right) = (q_2,(1\;3)(-)), \\
&\f{A}\left(q_2,????\right) = (q_{\rm f}, \turn_{\{3\}}(-)).
\end{align*}
In this case, the card-sequence $\crd{1}~\crd{2}~\crd{3}~\crd{4}$ is changed by the protocol $\mathcal{P}$ as follows:
\[ \crd{1}~\crd{2}~\crd{3}~\crd{4}~ \longrightarrow \overset{1}{\back}~\overset{2}{\back}~\overset{3}{\back}~\overset{4}{\back}~ \longrightarrow\overset{3}{\back}~\overset{2}{\back}~\overset{1}{\back}~\overset{4}{\back}~ \longrightarrow \overset{3}{\back}~\overset{2}{\back}~\crd{1}~\overset{4}{\back}~.\]
Thus, the final sequence is $ \back~\back~\crd{1}~\back~$.
\end{example}


\subsubsection{Shuffle protocols}\label{sss:reduction2}

A shuffle protocol\footnote{\md{Koch and Walzer \cite{KochFUN21} considered a similar notion and proposed a protocol for any uniform closed shuffles. The main difference of their model and our model is that their model allows a randomness generation in the head (see Section \ref{relatedworks} in Introduction). }} is a card-based protocol realizing a shuffle operation. 
It takes a card-sequence $\f{x} = (\f{x}_1, \f{x}_2, \ldots, \f{x}_n)$ such that $\f{back}(\f{x}) = (?, ?, \ldots, ?)$ as input and outputs $\f{y} = (\f{y}_1, \f{y}_2, \ldots, \f{y}_n)$ such that $\f{y} = \sigma(\f{x})$ for a permutation $\sigma$ is chosen from $\mathfrak{S}_n$ depending on some probability distribution:
\[
\underbrace{\back \, \back \, \cdots \, \back}_{\f{x}} ~ \underbrace{\crd{$h_1$} \, \crd{$h_2$} \, \cdots \, \crd{$h_k$}}_{\f{h}} ~ 
\lra~
\underbrace{\back \, \back \, \cdots \, \back}_{\f{y}} ~ \underbrace{\crd{$h_1$} \, \crd{$h_2$} \, \cdots \, \crd{$h_k$}}_{\f{h}}~,
\]
where $\f{h}$ is a card-sequence of helping cards. 
\md{Informally speaking, the correctness requires $\mathsf{y}=\mathsf{shuffle}_{(\Pi,\mathcal{F})}(\mathsf{x})$ and the security requires that no one learns nothing about the chosen permutation $\sigma \in \Pi$.}

\begin{definition}
Let $D_{\f{inp}}, D_{\f{help}}$ be decks, $U_{\f{inp}}$ an input set from $D_{\f{inp}}$, and $\f{h} \in \f{Seq}^{D_{\f{help}}}$ a card-sequence from $D_{\f{help}}$. 
We define an input set $U$ from $D = D_{\f{inp}} \cup D_{\f{help}}$ by $U = \{(\f{x}, \f{h}) \mid \f{x} \in U_{\f{inp}}\}$. 
A card-based protocol $\mathcal{P}=(D, U,Q,\f{A})$ is said to be a \textit{shuffle protocol} if the following conditions are satisfied:
\begin{enumerate}
\item[(a)] $\mathcal{P}$ always terminates within a fixed number of steps, i.e., it is a finite-runtime protocol;
\item[(b)] for any input sequence $(\f{x}, \f{h}) \in U$ \md{and for any final sequence $\f{y} \in \f{Fin}(\mathcal{P})$ of the form $\f{y} = (\f{x'}, \f{h})$, there exists a permutation $\sigma \in \mathfrak{S}_{|D_{\f{inp}}|}$ such that $\f{x'} = \sigma(\f{x})$;}
\item[(c)] for any input sequence $(\f{x}, \f{h})\in U$, \md{any card contained in $\f{x}$ has not been turned} at any step of a protocol execution.
\end{enumerate}
\md{
We say that $\mathcal{P}$ \md{implements} a shuffle $\shuffle_{(\Pi,\F)}$ if every permutation $\sigma$ in (b) belongs to $\Pi$ and it is chosen according to the distribution $\F$.}
We say that $\mathcal{P}$ is secure if for any $\f{x} \in U_{\f{inp}}$, \md{a random variable of $\sigma$} is stochastically independent of the random variable of the visible sequence-trace of $\mathcal{P}$. 
\end{definition}

\section{Graph shuffle protocols}\label{s:graph}

In this section, we construct a card-based protocol called  the graph shuffle protocol for a directed graph.
First, we introduce a graph shuffle in Subsection \ref{ss:graph}. Second, we construct the graph shuffle protocol, which is a shuffle protocol for any graph shuffle in Subsection \ref{ss:graphprotocol}. We note that our protocol requires PSSs only.

\subsection{Graph shuffle}\label{ss:graph}
First, we recall some fundamentals from graph theory; for example, see \cite{CLZ}.

A directed graph is a quadruple $G=(V_G,E_G,s_G,t_G)$ consisting of two sets $V_G$, $E_G$ and two maps $s_G,t_G:E_G\to V_G$.
Each element of $V_G$ (resp. $E_G$) is called a vertex (resp. an \md{edge}). 
\md{Note that there might be two or more edges from $a$ to $b$ for some $a, b \in V_G$, that is, $G$ admits multiple edges.} 
For an \md{edge} $e\in E_G$, we call $s_G(e)$ (resp. $t_G(e)$) the source (resp. the target) of $e$. We will commonly write $a\xrightarrow{e}b$ or $e:a\to b$ to indicate that an \md{edge} $e$ has the source $a$ and the target $b$, and identify $e$ with a pair $(s_G(e),t_G(e))$.
A directed graph $G$ is finite if two sets $V_G$ and $E_G$ are finite sets. 
In \md{this paper}, a graph means a finite directed graph with $n$ vertices and $m$ \md{edges}.

Let $G$ be a graph.
For a vertex  $v\in V_G$, we define the following three functions:
\[ \f{in}(v)=|\{e\in E_G\mid v=t_G(e)\}|,\quad \f{out}(v)=|\{e\in E_G \mid v=s_G(e)\}|, \quad\text{and}\quad \f{deg}(v)=\f{in}(v)+\f{out}(v).\]
The number $\f{deg}(v)$ is called \textit{the degree} of $v$.
We set $\f{Deg}_G = \{\f{deg}(v) \mid v \in V_G\}$.

For graphs $G$ and $G'$, a pair $f=(f_0,f_1):G\to G'$ consisting of maps $f_0:V_G\to V_{G'}$ and $f_1:E_G\to E_{G'}$ is a morphism of graphs if $(f_0\times f_0)\circ(s_G\times t_G)=(s_{G'}\times t_{G'})\circ f_1$ holds. In addition, if $f_0$ and $f_1$ are bijective, $f$ is called  \textit{an isomorphism} of graphs. In this case, we say that $G$ and $G'$ are isomorphic as graphs. 
In other words, two graphs $G$ and $G'$ are isomorphic as graphs when $x\xrightarrow{e}y$ in $G$ exists if and only if $f_0(x)\xrightarrow{f_1(e)}f_0(y)$ exists in $G'$.
We denote by $\f{Iso}(G,G')$ the set of all isomorphisms from $G$ to $G'$, and $\f{Iso}_0(G,G')$ the set of all $f_0$ such that $(f_0, *) \in \f{Iso}(G,G')$.
For a graph $G$, an isomorphism from $G$ to itself is called \textit{an automorphism}.
We denote by $\mathsf{Aut}(G)$ the set of all automorphisms of $G$, and $\mathsf{Aut}_0(G)$ the set of all $f_0$ such that \md{there exists $(f_0, f_1) \in \mathsf{Aut}(G)$}. \md{Then it is obvious that $\mathsf{Aut}(G)$ is a group by the composition of maps.
Furthermore, the group structure of $\mathsf{Aut}(G)$ induces the group structure on $\mathsf{Aut}_0(G)$.}
Note that, if $G$ has no multiple \md{edges}, then an automorphism $f = (f_0, f_1)\in \mathsf{Aut}(G)$ is determined by $f_0$. 
In the case that $G$ is an undirected graph, one can transform $G$ into the following directed graph $\overset{\to}{G}$:
\[ \text{ $V_{\overset{\to}{G}}=V_G$,\quad $E_{\overset{\to}{G}}=\{i\to j, \ j\to i\mid (i,j) \in E_G\}$}. \]

\begin{definition}\label{def:graphshuffle}
Let $G$ be a graph. 
The uniform closed shuffle $\shuffle_{(\aut_0(G), \F)}$ is called  \textit{the graph shuffle} for $G$ over $n$ cards. (Recall that $G$ has $n$ vertices.)
\end{definition}

\subsection{Graph shuffle protocols}\label{ss:graphprotocol}

In this subsection, we construct a graph shuffle protocol, which is a shuffle protocol of the graph shuffle for a graph $G = (V_G,E_G,s_G,t_G)$. 
We set $V_G=\{1,2,\ldots, n\}$. 
Let $D_{\f{inp}}=\{x_1,x_2,\ldots, x_n\}$ be any deck, and $U_{\f{inp}}$ any input set from $D_{\f{inp}}$. 
We set a card-sequence $\f{h}$ of helping cards as follows:
\[
\f{h} = \crd{$\stext{1}$} \, \crd{$\stext{2}$} \, \crd{$\stext{3}$} ~\cdots ~\crd{$\stext{n}$}~
\overbrace{\crd{1} \, \cdots \, \crd{1}}^{\f{deg}(1)} ~ \overbrace{\crd{2} \, \cdots \, \crd{2}}^{\f{deg}(2)} ~ \overbrace{\crd{3} \, \cdots \, \crd{3}}^{\f{deg}(3)} ~ \cdots \cdots~ \overbrace{\crd{$n$} \, \cdots \, \crd{$n$}}^{\f{deg}(n)}~.
\]
Thus the deck of helping cards is $D_{\f{help}} = \{\stext{1}, \stext{2}, \ldots, \stext{n}, 1^{\f{deg}(1)}, 2^{\f{deg}(2)}, \ldots, n^{\f{deg}(n)}\}$, where the superscript denotes the number of the symbol in the deck $D_{\f{help}}$. 
The deck $D$ is the union of $D_{\f{inp}}$ and $D_{\f{help}}$ as multisets and it consists of $2(n+m)$ symbols. 

For an input card-sequence $\f{x}\in U_{\f{inp}}$, our protocol proceeds as follows:

\begin{enumerate}
\item[(1)] Place the cards as follows.
\[
\underbrace{\back \, \back \, \cdots \, \back}_{\f{x}} ~ 
\underbrace{\crd{$\stext{1}$} \, \crd{$\stext{2}$} \, \crd{$\stext{3}$} ~\cdots ~\crd{$\stext{n}$}~
\crd{1} \, \cdots \, \crd{1} ~ \crd{2} \, \cdots \, \crd{2} ~ \crd{3} \, \cdots \, \crd{3} ~ \cdots \cdots~ \crd{$n$} \, \cdots \, \crd{$n$}}_{\f{h}}~. 
\]
\item[(2)] For each $i$, we define $\f{pile}[i]$ by
\[
\f{pile}[i]= \biggl( \,\dfrac{?}{\stext{i}}, \overbrace{\dfrac{?}{i}, \ldots, \dfrac{?}{i}}^{\f{deg}(i)} \,\biggr)  = \underset{\stext{i}}{\back}\, \overbrace{\underset{i}{\back} \, \cdots \, \underset{i}{\back}}^{\f{deg}(i)}.
\]
Arrange the card-sequence as $(\f{x}, \f{pile}[1], \f{pile}[2], \f{pile}[3], \ldots, \f{pile}[n])$, that is:
\[
\underbrace{\back \, \back \, \cdots \, \back}_{\f{x}} ~ \underbrace{\back \, \back \, \cdots \, \back}_{\f{pile}[1]}~ 
\underbrace{\back \, \back \, \cdots \, \back}_{\f{pile}[2]}~ 
\underbrace{\back \, \back \, \cdots \, \back}_{\f{pile}[3]}~\cdots\cdots~ 
\underbrace{\back \, \back \, \cdots \, \back}_{\f{pile}[n]}.
\]

\item[(3)]  For each $d \in \f{Deg}_G$, we set $V_G^{(d)} = \{v_1^{(d)}, v_2^{(d)}, \ldots, v_{\ell_d}^{(d)}\}$ for all vertices with degree $d$, and apply $\f{PSS}_{(\ell_d, d+1)}$ to the card-sequence  $(\f{pile}[v_1^{(d)}], \f{pile}[v_2^{(d)}], \ldots, \f{pile}[v_{\ell_d}^{(d)}])$.
Then we obtain a card-sequence
\[
\underbrace{\back \, \back \, \cdots \, \back}_{\f{x}} ~ \underbrace{\back \, \back \, \cdots \, \back}_{\f{pile}[\alpha_1]}~ \underbrace{\back \, \back \, \cdots \, \back}_{\f{pile}[\alpha_2]}~ \underbrace{\back \, \back \, \cdots \, \back}_{\f{pile}[\alpha_3]}~\cdots\cdots~ \underbrace{\back \, \back \, \cdots \, \back}_{\f{pile}[\alpha_n]}~.
\]
Let $\sigma \in \mathfrak{S}_n$ be the chosen permutation such that $\alpha_i = \sigma^{-1}(i)$. 

\item[(4)] For each $i\in V_G$ and $j \ra k \in E_G$, we set $\f{vertex}[i]= \left(\dfrac{?}{\stext{\alpha_i}}, \f{x}_i\right)$ and $\f{\md{edge}}[j \ra k]= \left(\dfrac{?}{\alpha_j}, \dfrac{?}{\alpha_k}\right)$, respectively.
Arrange the card-sequence\footnote{\md{Note that this rearrangement is possible without looking under the cards since the subscripts of $\alpha_i$ are public information.}} as follows:
\[
\underbrace{\back\,\back}_{\f{vertex}[1]}\,\underbrace{\back\,\back}_{\f{vertex}[2]}\,\cdots\,\underbrace{\back\,\back}_{\f{vertex}[n]}~~
\underbrace{\back\,\back}_{\f{\md{edge}}[e_1]}\,\underbrace{\back\,\back}_{\f{\md{edge}}[e_2]}\,\cdots\,\underbrace{\back\,\back}_{\f{\md{edge}}[e_m]}~,
\]
where $E_G = \{e_1, e_2, \ldots, e_m\}$. 

\item[(5)] Apply $\f{PSS}_{(m+n,2)}$ to the card-sequence as follows: 
\[
\bigg|\,\underbrace{\back\,\back}_{\f{vertex}[1]}\,\bigg|\,\underbrace{\back\,\back}_{\f{vertex}[2]}\,\bigg|\,\cdots\bigg|\,\underbrace{\back\,\back}_{\f{\md{edge}}[e_m]}\,\bigg|
~\ra~
\back\,\back~\back\,\back~\cdots~\back\,\back~\back\,\back\,.
\]
\item[(6)] For each pile, turn over the left card, and if it is a black-card, turn over the right card. 
Then sort $n+m$ piles\footnote{\md{It is not essential the order of pairs of helping cards.}} so that the left card is in ascending order via $\preccurlyeq$ as follows:
\[
\crd{$\stext{1}$}\,\back~\crd{$\stext{2}$}\,\back~\crd{$\stext{3}$}\,\back~\cdots~\crd{$\stext{n}$}\,\back~\crd{$i_1$}\,\crd{$j_1$}~\crd{$i_2$}\,\crd{$j_2$}~\crd{$i_3$}\,\crd{$j_3$}~\cdots~\crd{$i_m$}\,\crd{$j_m$}\,,
\]
where $i_1\preccurlyeq i_2 \preccurlyeq i_3 \preccurlyeq \cdots \preccurlyeq i_m$. 

\item[(7)] We define a graph $G'$ by $V_{G'} = V_G$ and $E_{G'} = \{i_1\ra j_1, i_2\ra j_2, i_3\ra j_3, \ldots, i_m\ra j_m\}$.

\item[(8)] Take an isomorphism $\psi:G\to G'$, and set $\beta_i := \psi^{-1}_0(i)$. 
Let $\f{y}_i$ be the right next card of $\crd{$\stext{\beta_i}$}\,$ and $\f{y} =(\f{y}_1, \f{y}_2, \ldots, \f{y}_n)$. 
Arrange the card-sequence as follows:
\[
\underbrace{\back \, \back \, \cdots \, \back}_{\f{y}} ~ 
\underbrace{\crd{$\stext{1}$} \, \crd{$\stext{2}$} \, \crd{$\stext{3}$} ~\cdots ~\crd{$\stext{n}$}~
\crd{1} \, \cdots \, \crd{1} ~ \crd{2} \, \cdots \, \crd{2} ~ \crd{3} \, \cdots \, \crd{3} ~ \cdots \cdots~ \crd{$n$} \, \cdots \, \crd{$n$}}_{\f{h}}~. 
\]
The output card-sequence for the input $\f{x}$ is $\f{y}$. 
\end{enumerate}

\begin{remark}
Regarding the number of cards, the number of cards in the proposed protocol is $2n + 2m$, of which $n + 2m$ are helping cards. 
As for the number of shuffles, it is \md{$|\f{Deg}_G|+1$}, and all of them are PSSs. 
We remark that the PSSs in Step (3) can be executed in parallel. 
\end{remark}

\begin{remark}\label{rem:iso}
In \md{Step (8)}, given two isomorphic graphs $G$ and $G'$, we need to solve the problem of finding one specific isomorphism between them. 
However, no polynomial-time algorithm for this problem has been found so far in general.
On the other hand, there exist polynomial-time algorithms to find isomorphisms for some \md{specific} graph classes. In addition, for small \md{specific} examples, an isomorphism can be computed by using a mathematical library for graph computation (e.g., Nauty \cite{Nauty}).  
\end{remark}

\subsection{Proof of correctness}

Let $\f{x} = (\f{x}_1, \f{x}_2, \ldots, \f{x}_n)$ be an input sequence and $\f{y} = (\f{y}_1, \f{y}_2, \ldots, \f{y}_n)$ a random variable of an output sequence of the protocol when $\f{x}$ is given as input. 
Fix a graph $G'$, which is defined in Step (7) following the opened result in Step (6). 
Let $\sigma \in \mathfrak{S}_n$ be a random variable of the permutation chosen by the PSSs in Step (3) such that $\alpha_i = \sigma^{-1}(i)$ for all $i \in V_G$. 
Since an \md{edge} $i \ra j \in E_G$ of $G$ corresponds to an \md{edge} $\alpha_i \ra \alpha_j = \sigma^{-1}(i) \ra \sigma^{-1}(j) \in E_{G'}$ of $G'$, there is an isomorphism $\phi = (\phi_0, \phi_1) \in \f{Iso}(G, G')$ such that $\phi_0 = \sigma^{-1}$. 
From the property of the PSSs, the permutation $\phi_0$ is a uniform random variable on $\f{Iso}_0(G, G')$. 
Let $\psi = (\psi_0, \psi_1) \in \f{Iso}(G', G)$ be a random variable of the isomorphism chosen in Step (9). 

We first claim that $\f{y} = \psi_0 \circ \phi_0 (\f{x})$. 
This is shown by observing a sequence of red cards $\crd{$\stext{1}$} \, \crd{$\stext{2}$} \, \crd{$\stext{3}$} \, \cdots \,\crd{$\stext{n}$}\,$. 
Hereafter, for the sake of clarity, we do not distinguish face-up $\dfrac{\stext{i}}{?}$ and face-down $\dfrac{?}{\stext{i}}$ and use ``$\stext{i}$" to denote the red card $i$. 
In Step (1), the sequence of red cards omitting other cards is $(\stext{1}, \stext{2}, \ldots, \stext{n})$. 
In Steps (3), (6), and (8), it is arranged as follows:
\[
\underset{\text{Step (1)}}{(\stext{1}, \stext{2}, \ldots, \stext{n})}
\xra{\phi_0^{-1}}
\underset{\text{Step (3)}}{(\stext{\alpha_1}, \stext{\alpha_2}, \ldots, \stext{\alpha_n})}
\xra{\phi_0}
\underset{\text{Step (6)}}{(\stext{1}, \stext{2}, \ldots, \stext{n})}
\xra{\psi_0}
\underset{\text{Step (8)}}{(\stext{\beta_1}, \stext{\beta_2}, \ldots, \stext{\beta_n})}.
\]
Since the input sequence $\f{x}$ is arranged as $((\stext{\alpha_1}, \f{x}_1), (\stext{\alpha_2}, \f{x}_2), \ldots, (\stext{\alpha_n}, \f{x}_n))$ in Step (3), the permutation $\psi_0 \circ \phi_0$ is applied to $\f{x}$. 
Thus, it holds $\f{y} = \psi_0 \circ \phi_0 (\f{x})$. 
We note that $\psi \circ \phi_0$ is an automorphism of $G$.  


It remains to prove that the distribution of $\psi_0\circ\phi_0 \in \aut_0(G)$ is uniformly random. 
We note that given the graph $G'$, the distributions of $\phi_0$ and $\psi_0$ are independent. 
This is because the choice of $\psi_0$ depends on the opened symbols in Step (6) only, and they are independent of $\phi_0$ due to the PSS in Step (5). 
Thus, we can change the order of choice without harming the distributions of $\phi_0, \psi_0$: first, $\psi_0$ is chosen, and then $\phi_0$ is chosen. 
Since the distribution of $\phi_0 \in \f{Iso}_0(G, G')$ is uniformly random, it is sufficient to show that the function
\[ 
\begin{array}{cccc}
\Phi:  & \f{Iso}_0(G, G') &  \longrightarrow &  \aut_0(G) \\
           & \phi_0                                                    & \longmapsto & \psi_0 \circ\phi_0
 \end{array}
 \]
is bijective. 

We first prove that $\Phi$ is injective. 
Suppose that $\Phi(\phi'_0) = \Phi(\phi''_0)$ for some $\phi'_0, \phi''_0 \in \f{Iso}_0(G, G')$, that is,  $\psi_0 \circ\phi'_0 = \psi_0 \circ\phi''_0$. 
Since $\psi_0$ is a bijection, $\phi'_0 = \phi''_0$ holds.
Thus $\Phi$ is injective. 
We next prove that $\Phi$ is surjective. 
For any $\tau\in\aut_0(G)$, we have 
\[ \tau = \psi_0\circ\psi_0^{-1}\circ \tau = \Phi (\psi_0^{-1}\circ \tau). \] 
It yields that $\Phi$ is surjective. 
Therefore, $\Phi$ is bijective. 

This shows that the distribution of $\psi \circ \sigma^{-1}$ is uniformly random, and hence our protocol is correct. 


\subsection{Proof of security}

In the proof of the correctness, we have already claimed that the distribution of the opened symbols in Step (6) is independent of $\sigma$ due to the PSS in Step (5). 
Since cards are opened in Step (6) only, this shows a distribution of the permutation $\psi\circ \sigma^{-1} \in \aut(G)$ is independent of the distribution of the visible sequence-trace of our protocol. 
Therefore, our protocol is secure. 

\subsection{Example of our protocol for a graph}
Let $G$ be a directed graph with $5$ vertices as follows:
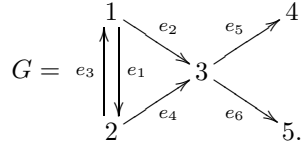
\[ G=  \begin{xy}
                     (0,8)*[o]+{1}="1",(0,-8)*[o]+{2}="2",(12,0)*[o]+{3}="3",(24,8)*[o]+{4}="4",(24,-8)*[o]+{5\md{.}}="5",
                     \ar @<1mm>"1";"2"^{e_1}
                     \ar @<1mm>"2";"1"^{e_3}
                     \ar "1";"3"^{e_2}
                     \ar "2";"3"_{e_4}
                     \ar "3";"4"^{e_5}
                     \ar "3";"5"_{e_6}
            \end{xy} \]
We perform our graph shuffle protocol for $G$. 
Let $D_{\f{inp}}$ be an arbitrary deck with $D_{\f{inp}} = \{x_1, x_2, x_3, x_4, x_5\}$. 
The card-sequence $\f{h}$ of helping cards is defined as follows:
\[
\f{h} = \crd{$\stext{1}$} \, \crd{$\stext{2}$} \, \crd{$\stext{3}$} \, \crd{$\stext{4}$} \, \crd{$\stext{5}$}~
\crd{1} \, \crd{1} \, \crd{1}  ~ \crd{2} \, \crd{2} \, \crd{2} ~ \crd{3} \, \crd{3} \, \crd{3} \, \crd{3} ~  \crd{4} ~ \crd{5}\,\md{.}
\]
Set $D_{\f{help}} = \{\stext{1}, \stext{2}, \stext{3}, \stext{4}, \stext{5},1, 1, 1, 2, 2, 2, 3, 3, 3, 3, 4, 5\}$ and $D=D_{\f{inp}}\cup D_{\f{help}}$. 
For an input card-sequence $\f{x} = (\f{x}_1, \f{x}_2, \f{x}_3, \f{x}_4, \f{x}_5) \in U$, the graph shuffle protocol proceeds as follows:

\begin{enumerate}
\item[(1)] Place the cards such as:
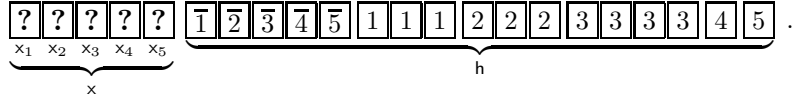
\[
\underbrace{\underset{\f{x}_1}{\back} \, \underset{\f{x}_2}{\back} \, \underset{\f{x}_3}{\back} \, \underset{\f{x}_4}{\back} \, \underset{\f{x}_5}{\back}}_{\f{x}} ~ 
\underbrace{\crd{$\stext{1}$} \, \crd{$\stext{2}$} \, \crd{$\stext{3}$} \, \crd{$\stext{4}$} \, \crd{$\stext{5}$}~
\crd{1} \, \crd{1} \, \crd{1}  ~ \crd{2} \, \crd{2} \, \crd{2} ~ \crd{3} \, \crd{3} \, \crd{3} \, \crd{3} ~  \crd{4} ~ \crd{5}}_{\f{h}}~. 
\]
\item[(2)] Arrange the card-sequence as follows:
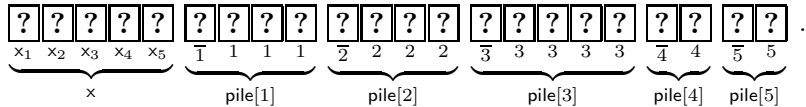
\[
\underbrace{\underset{\f{x}_1}{\back} \, \underset{\f{x}_2}{\back} \, \underset{\f{x}_3}{\back} \, \underset{\f{x}_4}{\back} \, \underset{\f{x}_5}{\back}}_{\f{x}} ~ \underbrace{\underset{\stext{1}}{\back} \, \underset{1}{\back} \, \underset{1}{\back} \, \underset{1}{\back}}_{\f{pile}[1]}~ 
\underbrace{\underset{\stext{2}}{\back} \, \underset{2}{\back} \, \underset{2}{\back} \, \underset{2}{\back}}_{\f{pile}[2]}~ 
\underbrace{\underset{\stext{3}}{\back} \, \underset{3}{\back} \, \underset{3}{\back} \, \underset{3}{\back} \, \underset{3}{\back}}_{\f{pile}[3]}~ 
\underbrace{\underset{\stext{4}}{\back} \, \underset{4}{\back}}_{\f{pile}[4]}~ 
\underbrace{\underset{\stext{5}}{\back} \, \underset{5}{\back}}_{\f{pile}[5]}~.
\]

\item[(3)] Perform $\f{PSS}_{(2, 4)}$ and $\f{PSS}_{(2, 2)}$ as follows:
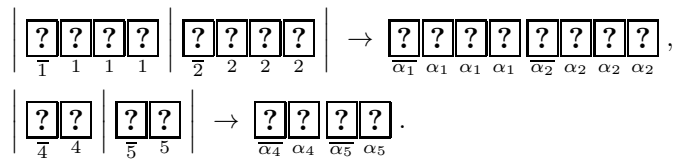
\begin{align*}
&\bigg|~
\underset{\stext{1}}{\back} \, \underset{1}{\back} \, \underset{1}{\back} \, \underset{1}{\back}
~\bigg|~
\underset{\stext{2}}{\back} \, \underset{2}{\back} \, \underset{2}{\back} \, \underset{2}{\back}
~\bigg|
~~\ra~~
\underset{\stext{\alpha_1}}{\back} \,\underset{\alpha_1}{\back} \, \underset{\alpha_1}{\back} \,\underset{\alpha_1}{\back}~~
\underset{\stext{\alpha_2}}{\back} \, \underset{\alpha_2}{\back} \, \underset{\alpha_2}{\back} \, \underset{\alpha_2}{\back}~,\\
&\bigg|~
\underset{\stext{4}}{\back} \, \underset{4}{\back}
~\bigg|~
\underset{\stext{5}}{\back} \, \underset{5}{\back}
~\bigg|
~~\ra~~
\underset{\stext{\alpha_4}}{\back} \, \underset{\alpha_4}{\back}~~
\underset{\stext{\alpha_5}}{\back} \, \underset{\alpha_5}{\back}~.
\end{align*}
By setting $\alpha_3 = 3$, we have the following card-sequence:
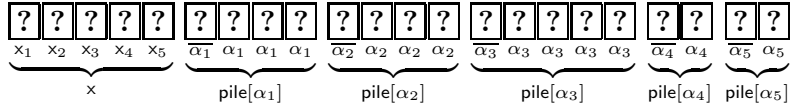
\[
\underbrace{\underset{\f{x}_1}{\back} \, \underset{\f{x}_2}{\back} \, \underset{\f{x}_3}{\back} \, \underset{\f{x}_4}{\back} \, \underset{\f{x}_5}{\back}}_{\f{x}} ~ \underbrace{\underset{\stext{\alpha_1}}{\back} \, \underset{\alpha_1}{\back} \, \underset{\alpha_1}{\back} \, \underset{\alpha_1}{\back}}_{\f{pile}[\alpha_1]}~ 
\underbrace{\underset{\stext{\alpha_2}}{\back} \, \underset{\alpha_2}{\back} \, \underset{\alpha_2}{\back} \, \underset{\alpha_2}{\back}}_{\f{pile}[\alpha_2]}~ 
\underbrace{\underset{\stext{\alpha_3}}{\back} \, \underset{\alpha_3}{\back} \, \underset{\alpha_3}{\back} \, \underset{\alpha_3}{\back} \, \underset{\alpha_3}{\back}}_{\f{pile}[\alpha_3]}~ 
\underbrace{\underset{\stext{\alpha_4}}{\back} \, \underset{\alpha_4}{\back}}_{\f{pile}[\alpha_4]}~ 
\underbrace{\underset{\stext{\alpha_5}}{\back} \, \underset{\alpha_5}{\back}}_{\f{pile}[\alpha_5]}~.
\]

\item[(4)] Arrange the card-sequence as follows:
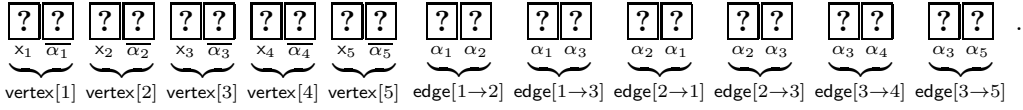
\[
\underbrace{\underset{\f{x}_1}{\back}\,\underset{\stext{\alpha_1}}{\back}}_{\f{vertex}[1]}\,
\underbrace{\underset{\f{x}_2}{\back}\,\underset{\stext{\alpha_2}}{\back}}_{\f{vertex}[2]}\,
\underbrace{\underset{\f{x}_3}{\back}\,\underset{\stext{\alpha_3}}{\back}}_{\f{vertex}[3]}\,
\underbrace{\underset{\f{x}_4}{\back}\,\underset{\stext{\alpha_4}}{\back}}_{\f{vertex}[4]}\,
\underbrace{\underset{\f{x}_5}{\back}\,\underset{\stext{\alpha_5}}{\back}}_{\f{vertex}[5]}~~
\underbrace{\underset{\alpha_1}{\back}\,\underset{\alpha_2}{\back}}_{\f{\md{edge}}[1\ra 2]}\,
\underbrace{\underset{\alpha_1}{\back}\,\underset{\alpha_3}{\back}}_{\f{\md{edge}}[1\ra 3]}\,
\underbrace{\underset{\alpha_2}{\back}\,\underset{\alpha_1}{\back}}_{\f{\md{edge}}[2\ra 1]}\,
\underbrace{\underset{\alpha_2}{\back}\,\underset{\alpha_3}{\back}}_{\f{\md{edge}}[2\ra 3]}\,
\underbrace{\underset{\alpha_3}{\back}\,\underset{\alpha_4}{\back}}_{\f{\md{edge}}[3\ra 4]}\,
\underbrace{\underset{\alpha_3}{\back}\,\underset{\alpha_5}{\back}}_{\f{\md{edge}}[3\ra 5]}\,.
\]

\item[(5)] Apply $\f{PSS}_{(11,2)}$ to the card-sequence as follows: 
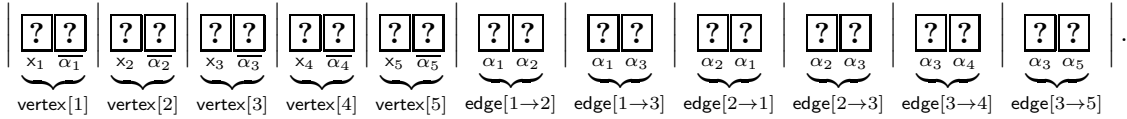
\[
\bigg|\underbrace{\underset{\f{x}_1}{\back}\,\underset{\stext{\alpha_1}}{\back}}_{\f{vertex}[1]}
\bigg|\underbrace{\underset{\f{x}_2}{\back}\,\underset{\stext{\alpha_2}}{\back}}_{\f{vertex}[2]}
\bigg|\underbrace{\underset{\f{x}_3}{\back}\,\underset{\stext{\alpha_3}}{\back}}_{\f{vertex}[3]}
\bigg|\underbrace{\underset{\f{x}_4}{\back}\,\underset{\stext{\alpha_4}}{\back}}_{\f{vertex}[4]}
\bigg|\underbrace{\underset{\f{x}_5}{\back}\,\underset{\stext{\alpha_5}}{\back}}_{\f{vertex}[5]}
\bigg|\underbrace{\underset{\alpha_1}{\back}\,\underset{\alpha_2}{\back}}_{\f{\md{edge}}[1\ra 2]}
\bigg|\underbrace{\underset{\alpha_1}{\back}\,\underset{\alpha_3}{\back}}_{\f{\md{edge}}[1\ra 3]}
\bigg|\underbrace{\underset{\alpha_2}{\back}\,\underset{\alpha_1}{\back}}_{\f{\md{edge}}[2\ra 1]}
\bigg|\underbrace{\underset{\alpha_2}{\back}\,\underset{\alpha_3}{\back}}_{\f{\md{edge}}[2\ra 3]}
\bigg|\underbrace{\underset{\alpha_3}{\back}\,\underset{\alpha_4}{\back}}_{\f{\md{edge}}[3\ra 4]}
\bigg|\underbrace{\underset{\alpha_3}{\back}\,\underset{\alpha_5}{\back}}_{\f{\md{edge}}[3\ra 5]}\bigg|\,.
\]

\item[(6)] For each pile, turn over the left card, and if it is a black-card, turn over the right card. 
The following card-sequence is an example outcome:
\[
\crd{$\stext{5}$}\,\back~
\crd{1}\,\crd{3}~
\crd{2}\,\crd{3}~
\crd{$\stext{4}$}\,\back~
\crd{$\stext{2}$}\,\back~
\crd{2}\,\crd{1}~
\crd{$\stext{1}$}\,\back~
\crd{3}\,\crd{5}~
\crd{3}\,\crd{4}~
\crd{1}\,\crd{2}~
\crd{$\stext{3}$}\,\back\,.
\]
Sort $11$ piles so that the left card is in ascending order via $\preccurlyeq$ as follows:
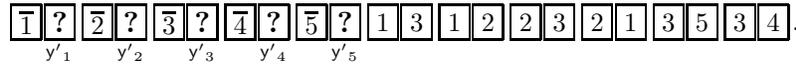
\[
\crd{$\stext{1}$}\,\underset{\f{y'}_1}{\back}~\crd{$\stext{2}$}\,\underset{\f{y'}_2}{\back}~\crd{$\stext{3}$}\,\underset{\f{y'}_3}{\back}~\crd{$\stext{4}$}\,\underset{\f{y'}_4}{\back}~\crd{$\stext{5}$}\,\underset{\f{y'}_5}{\back}~
\crd{1}\,\crd{3}~\crd{1}\,\crd{2}~\crd{2}\,\crd{3}~\crd{2}\,\crd{1}~\crd{3}\,\crd{5}~\crd{3}\,\crd{4}\,.~
\]

\item[(7)] Define a graph $G'$ by $V_{G'} = \{1, 2, 3, 4, 5\}$ and $E_{G'} = \{1\ra3, 1\ra2, 2\ra3, 2\ra1, 3\ra4, 3\ra5\}$;
\[ G'=  \begin{xy}
                     (0,8)*[o]+{1}="1",(0,-8)*[o]+{2}="2",(12,0)*[o]+{3}="3",(24,8)*[o]+{4}="4",(24,-8)*[o]+{5.}="5",
                     \ar @<1mm>"1";"2"^{}
                     \ar @<1mm>"2";"1"^{}
                     \ar "1";"3"^{}
                     \ar "2";"3"_{}
                     \ar "3";"4"^{}
                     \ar "3";"5"_{}
            \end{xy}\]

\item[(8)] Take an isomorphism $\psi: G\to G'$ defined by 
\[
1\longmapsto 2,\quad 2\longmapsto 1,\quad 3\longmapsto 3 ,\quad 4\longmapsto 4,\quad 5\longmapsto 5.
\]
Arrange the above card-sequence as follows:
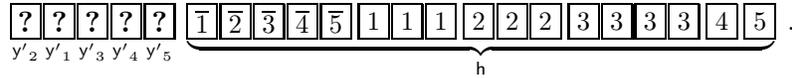
\[
\underset{\f{y'}_2}{\back} \, \underset{\f{y'}_1}{\back} \, \underset{\f{y'}_3}{\back} \, \underset{\f{y'}_4}{\back} \, \underset{\f{y'}_5}{\back}~ 
\underbrace{\crd{$\stext{1}$} \, \crd{$\stext{2}$} \, \crd{$\stext{3}$} \, \crd{$\stext{4}$} \, \crd{$\stext{5}$}~
\crd{1} \, \crd{1} \, \crd{1}  ~ \crd{2} \, \crd{2} \, \crd{2} ~ \crd{3} \, \crd{3} \, \crd{3} \, \crd{3} ~  \crd{4} ~ \crd{5}}_{\f{h}}~. 
\]
The output card-sequence for the input $\f{x}$ is $(\f{y'}_2, \f{y'}_1, \f{y'}_3, \f{y'}_4, \f{y'}_5)$. 
\end{enumerate}

\subsection{Implication of our protocol}\label{ss:implication}

\md{In this subsection, we consider several interesting graph shuffles.}

\md{
We first observe that a RC for $n$ cards are graph shuffles for the directed $n$-cycle graph $\overset{\ra}{C_n}$ (see Section \ref{4-1}). 
Since it holds $2n + 2m = 4n$ and $|\f{Deg}_G|+1 = 2$, a RC can be done by $4n$ cards and two PSSs. 
In Section \ref{4-1}, the number of cards is improved to $3n$.
We remark that our graph shuffle protocol works even for a sequence of piles each having equivalent number of face-down cards. 
Thus a pile-shifting shuffle (i.e., a pile-version of RC) can be done by the same number of helping cards. 
In particular, for a pile-shifting shuffle for $n$ piles of $m$ cards, it can be done by $nm + 3n$ cards and two PSSs. 
We note that PSSs and RBCs are graph shuffles for graphs with no edges in this sense.}

\md{A graph shuffle for the undirected $n$-cycle graph $C_n$ is equivalent to the \emph{dihedral shuffle}, which is introduced by Niemi and Renvall \cite{Niemi98}. 
Since it holds $2n + 2m = 5n$ and $|\f{Deg}_G|+1 = 2$, our result implies that a RC can be done by $5n$ cards and two PSSs. 
In Section \ref{sec:dihedral}, the number of cards is improved to $3n$, although the number of PSSs is increased to three.}

\md{For a cyclic group $\Pi = \langle (1\;2)(3\;4\;5\;6) \rangle$, a uniform closed shuffle $(\shuffle, \Pi, \F)$ is a graph shuffle for $G$ where $V_G = \{1, 2, 3, 4, 5, 6\}$ and $E_G = E_1 \cup E_2 \cup E_3$ with $E_1 = \{1\to 2, 2 \to 1\}$, $E_2 = \{3\to 4, 4\to 5, 5\to 6, 6\to 4\}$, and $E_3 = \{1 \to 3, 1\to 5, 2\to 4, 2 \to 6\}$. 
Since it holds $\mathsf{Aut}_0(G) =  \langle (1\;2)(3\;4\;5\;6) \rangle$, we can conclude that a graph shuffle for $G$ is equivalent to a uniform closed shuffle $(\shuffle, \Pi, \F)$. 
Since it holds $2n + 2m = 32$ and $|\f{Deg}_G|+1 = 3$, our result implies that it can be done by $32$ cards and three PSSs. 
By generalizing this idea, for any cyclic group $\Pi = \langle \pi \rangle$, a uniform closed shuffle $(\shuffle, \Pi, \F)$ is a graph shuffle for some graph. 
}

\section{Efficiency improvements for graph shuffles for cycles}\label{s:app}

In this section, we \md{implement} efficient graph shuffle protocols for some \md{specific} graph classes. 
In particular, we improve the number of cards in our protocol.  

\subsection{The $n$-cycle graph} \label{4-1}
First, we consider the $n$-cycle graph $\overset{\ra}{C_n}$:
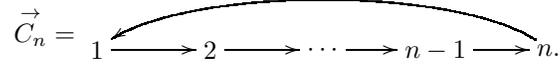
\[ \overset{\ra}{C_n}= \begin{xy}
                     (0,-2)*[o]+{1}="1",(15,-2)*[o]+{2}="2",(30,-2)*[o]+{\cdots}="3",(45,-2)*[o]+{n-1}="4",(60,-2)*[o]+{n.}="n",
                     \ar "1";"2"^{}
                     \ar "2";"3"^{}
                     \ar "3";"4"_{}
                     \ar "4";"n"_{}
                     \ar @(lu,ur)"n";"1"_{}
            \end{xy}\]
The graph shuffle for  $\overset{\ra}{C_n}$ is equivalent to a RC of $n$ cards since the automorphism group $\aut(\overset{\ra}{C_n})$ is isomorphic to the cyclic group of degree $n$.
If we apply our graph shuffle protocol for $\overset{\ra}{C_n}$ proposed in Section 3, we need $4n$ cards.
In this subsection, we propose a graph shuffle protocol for $\overset{\ra}{C_n}$ with $3n$ cards only. 

\md{Before describing the improved protocol, we shortly mention how to improve the number of cards. The idea\footnote{We remark that this idea works for every graphs such that all vertices have the same degree.} is to remove the red cards by making a pile of $(\f{x}_i, \alpha_i, \alpha_{i+1})$ instead of a pile of $(\f{x}_i, \stext{\alpha_i})$ and a pile of $(\alpha_i, \alpha_{i+1})$ in the previous protocol. Since all vertices of $\overset{\ra}{C_n}$ have the same degree, all piles of $(\f{x}_i, \alpha_i, \alpha_{i+1})$ have the same number of cards and thus the final randomization (corresponding to Step (5) in the previous protocol) can be done by a single PSS.}


Let $D_{\f{inp}}=\{x_1,x_2,\ldots, x_n\}$ be an arbitrary deck and $D_{\f{help}} = \{1, 1, 2, 2, 3, 3, \ldots, n, n\}$ a deck of the symbols of $2n$ helping cards. 
The sequence of helping cards $\f{h}$ is defined as follows:
\[ 
\f{h} = \crd{1}\,\crd{1}~\crd{2}\,\crd{2}~\crd{3}\,\crd{3}~\cdots\crd{$n$}\,\crd{$n$}~. \]
For $i=1,2,\ldots ,n$, we set $\f{pile}[i]=\left(\dfrac{?}{i}, \dfrac{?}{i}\right)$.

\begin{enumerate}
\item[(1)] Place the $3n$ cards as follows:
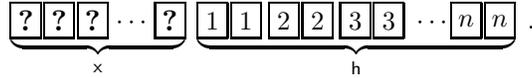
\[
\underbrace{\back \, \back \, \back \, \cdots \, \back}_{\f{x}} ~ 
\underbrace{\crd{1}\,\crd{1}~\crd{2}\,\crd{2}~\crd{3}\,\crd{3}~\cdots\crd{$n$}\,\crd{$n$}}_{\f{h}}~. 
\]

\item[(2)] Arrange the card-sequence as follows:
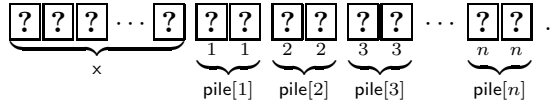
\[
\underbrace{\back \, \back \, \back \, \cdots \, \back}_{\f{x}} ~ 
\underbrace{\underset{1}{\back} \, \underset{1}{\back}}_{\f{pile}[1]}~ 
\underbrace{\underset{2}{\back} \, \underset{2}{\back}}_{\f{pile}[2]}~ 
\underbrace{\underset{3}{\back} \, \underset{3}{\back}}_{\f{pile}[3]}~ 
~\cdots~ 
\underbrace{\underset{n}{\back} \, \underset{n}{\back}}_{\f{pile}[n]}~ .
\]
Apply $\PSS_{(n,2)}$ to $(\f{pile}[1], \f{pile}[2] ,\ldots,\f{pile}[n])$ and then we obtain the card-sequence as follows:
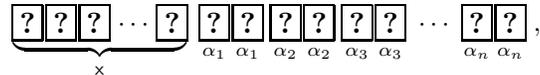
\[
\underbrace{\back \, \back \, \back \, \cdots \, \back}_{\f{x}} ~ 
\underset{\alpha_1}{\back} \, \underset{\alpha_1}{\back}
~
\underset{\alpha_2}{\back} \, \underset{\alpha_2}{\back}
~
\underset{\alpha_3}{\back} \, \underset{\alpha_3}{\back}
~
\cdots
~
\underset{\alpha_n}{\back} \, \underset{\alpha_n}{\back}~,
\]
where $\{\alpha_1,\alpha_2,\ldots,\alpha_n\}=\{1,2,\ldots, n\}$.

\item[(3)] Arrange the card-sequence as follows:
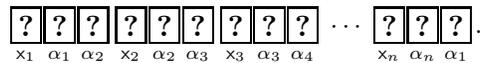
\[ 
\underset{\f{x}_1}{\back} \,
\underset{\alpha_1}{\back} \, \underset{\alpha_2}{\back} ~
\underset{\f{x}_2}{\back} \,
\underset{\alpha_2}{\back} \, \underset{\alpha_3}{\back} ~
\underset{\f{x}_3}{\back} \,
\underset{\alpha_3}{\back} \, \underset{\alpha_4}{\back} ~
\cdots ~
\underset{\f{x}_n}{\back} \,
\underset{\alpha_n}{\back} \, \underset{\alpha_1}{\back}\,.
\]

\item[(4)] Apply $\PSS_{(n,3)}$ to the card-sequence as follows:
\[
\begin{tabular}{|c|c|c|c|c|}
$\back\, \back\, \back$ &
$\back\, \back\, \back$ &
$\back\, \back\, \back$ &
$\cdots$ &
$\back\, \back\, \back$
\end{tabular}\,.
\]

\item[(5)] For all piles, turn over the second and third cards. Let $a_i, b_i \in \{1,2,\ldots, n\}$ be the opened symbols of the second and third cards, respectively, in the $i$-th pile as follows:
\[
\back\,\crd{$a_1$}\,\crd{$b_1$}~~\back\,\crd{$a_2$}\,\crd{$b_2$}~~\back\,\crd{$a_3$}\,\crd{$b_3$}~~\cdots~~\back\,\crd{$a_n$}\,\crd{$b_n$}~\md{.}
\]

\item[(6)] Arrange $n$ piles so that $(c_1, d_1) = (a_1, b_1)$, $d_i = c_{i+1}$, $(1 \leq i \leq n-1)$, and $d_n = c_1$ as follows:
\[
\underset{\f{y}_1}{\back} \,\crd{$c_1$}\,\crd{$d_1$}~~\underset{\f{y}_2}{\back} \,\crd{$c_2$}\,\crd{$d_2$}~~\underset{\f{y}_3}{\back} \,\crd{$c_3$}\,\crd{$d_3$}~~\cdots~~\underset{\f{y}_n}{\back} \,\crd{$c_n$}\,\crd{$d_n$}~.
\]
After that, we arrange the card-sequence as follows:
\[
\underset{\f{y}_1}{\back} \, \underset{\f{y}_2}{\back} \, \underset{\f{y}_3}{\back} \, \cdots \, \underset{\f{y}_n}{\back} ~ 
\underbrace{\crd{1}\,\crd{1}~\crd{2}\,\crd{2}~\crd{3}\,\crd{3}~\cdots\crd{$n$}\,\crd{$n$}}_{\f{h}}~. 
\]
Then the output card-sequence is $\mathsf{y}=(\f{y}_1,\f{y}_2,\ldots, \f{y}_n)$.
\end{enumerate}

\md{We show the correctness of the protocol. Let $\mathsf{x}=(\mathsf{x}_1,\ldots, \mathsf{x}_n)$ be an input sequence. Assume that the protocol outputs the sequence $\mathsf{y}=(\mathsf{y}_1,\ldots, \mathsf{y}_n)$ when $\mathsf{x}$ is given as input.
First, we see that $\mathsf{y}=\sigma (\mathsf{x})$ for some $\sigma$ in the cyclic group of degree $n$. 
For $i=1,\ldots, n$, we set $P_i=(\mathsf{x}_i,\alpha_i,\alpha_{i+1})$, where $\alpha_{n+1}=\alpha_1$, in Step (3) and put $P=(P_1,P_2,\ldots, P_n)$. 
Let $Q=(Q_1,\ldots, Q_n)=(P_{\sigma^{-1}(1)},\ldots, P_{\sigma^{-1}(n)})$ for some $\sigma\in\mathfrak{S}_n$. Then, $Q$ is obtained by $\sigma$ in the cyclic group of degree $n$ if and only if the third entry of $Q_i$ and the second entry of $Q_{i+1}$ are same for any $1\leq i\leq n-1$. It follows that the components of obtained sequence in Step (6) are sorted in a cyclic fashion of $P$. Therefore, $\mathsf{y}$ is equal to $\sigma (\mathsf{x})$ for some $\sigma$ in the cyclic group of degree $n$. 
Note that each element $\sigma$ of the cyclic group is determined by $\sigma^{-1}(1)$, and it is determined by $d_1$.  
For each $k\in\{1,2,\ldots, n\}$, the probability that $k=d_1$ is $\dfrac{1}{n}$ since $d_1$ is dependent on the PSS in Step (4) only. Thus the distribution of $\sigma$ is uniformly random, and hence the protocol is correct.}

\md{We show the security of the protocol. Assume that $\sigma\in\mathfrak{S}_n$ and $\tau\in \mathfrak{S}_n$ are chosen in Steps (2) and (4), respectively. Then the first card in the $i$-th pile in Step (5) is $\mathsf{x}_{\tau^{-1}(i)}$. On the other hand, the second and third cards in the $i$-th pile in Step (5) are $a_i=\tau^{-1}\sigma^{-1}(i)$ and $b_i=\tau^{-1}\sigma^{-1}(i+1)$. Here, we consider $n+1$ as $1$. 
This implies that these opened symbols $a_1,\ldots, a_n$ and $b_1,\ldots, b_n$ do not allow us to guess the first card of any pile since $\sigma$ is chosen uniformly at random in Step (4). Therefore, the protocol is secure.}

\subsection{The undirected $n$-cycle}\label{sec:dihedral}
Next, we consider the undirected $n$-cycle graph $C_n$:
\[ C_n= \begin{xy}
                     (0,-2)*[o]+{1}="1",(15,-2)*[o]+{2}="2",(30,-2)*[o]+{\cdots}="3",(45,-2)*[o]+{n-1}="4",(60,-2)*[o]+{n.}="n",
                     \ar @{-} "1";"2"^{}
                     \ar @{-} "2";"3"^{}
                     \ar @{-} "3";"4"_{}
                     \ar @{-} "4";"n"_{}
                     \ar @{-} @(lu,ur)"n";"1"_{}
            \end{xy}\]
\md{Recall that we regard undirected edge as two directed edges with opposite directions (see the paragraph just before Definition \ref{def:graphshuffle}).} 
The automorphism group $\aut(C_n)$ is isomorphic to the dihedral group of degree $n$. 
\md{For example, the graph shuffle for $C_n$ when $n = 4$ is given as follows: 
\[
\crd{1}~\crd{2}~\crd{3}~\crd{4}~ \longmapsto
\begin{cases}
~\crd{1}~\crd{2}~\crd{3}~\crd{4}~\\
~\crd{2}~\crd{3}~\crd{4}~\crd{1}~\\
~\crd{3}~\crd{4}~\crd{1}~\crd{2}~\\
~\crd{4}~\crd{1}~\crd{2}~\crd{3}~\\
~\crd{4}~\crd{3}~\crd{2}~\crd{1}~\\
~\crd{3}~\crd{2}~\crd{1}~\crd{4}~\\
~\crd{2}~\crd{1}~\crd{4}~\crd{3}~\\
~\crd{1}~\crd{4}~\crd{3}~\crd{2}~,
\end{cases}
\]
where each sequence is obtained with probability $1/8$.}
If we apply the graph shuffle for $C_n$, we need $6n$ cards.
In this subsection, we propose a graph shuffle protocol for $C_n$ with $3n$ cards only. 

\md{
For an undirected $n$-cyclic graph, even though it has $2n$ edges, the number of cards corresponding to edges is reduced to $n$ pairs of cards using the symmetry of the graph.
This improvement is done by the pile-scramble shuffle in Step (3) in the below protocol. }

Let $D_{\f{inp}}=\{x_1,x_2,\ldots, x_n\}$ be an arbitrary deck and $D_{\f{help}} = \{1, 1, 2, 2, 3, 3, \ldots, n, n\}$ a deck of the symbols of $2n$ helping cards. 
The sequence of helping cards $\f{h}$ is defined as follows:
\[ 
\f{h} = \crd{1}\,\crd{1}~\crd{2}\,\crd{2}~\crd{3}\,\crd{3}~\cdots\crd{$n$}\,\crd{$n$}~. \]
For $i=1,2,\ldots ,n$, we set $\f{pile}[i]=\left(\dfrac{?}{i}, \dfrac{?}{i}\right)$.

\begin{enumerate}
\item[(1)] Place the $3n$ cards as follows:
\[
\underbrace{\back \, \back \, \back \, \cdots \, \back}_{\f{x}} ~ 
\underbrace{\crd{1}\,\crd{1}~\crd{2}\,\crd{2}~\crd{3}\,\crd{3}~\cdots\crd{$n$}\,\crd{$n$}}_{\f{h}}~. 
\]

\item[(2)] Arrange the card-sequence as follows:
\[
\underbrace{\back \, \back \, \back \, \cdots \, \back}_{\f{x}} ~ 
\underbrace{\underset{1}{\back} \, \underset{1}{\back}}_{\f{pile}[1]}~ 
\underbrace{\underset{2}{\back} \, \underset{2}{\back}}_{\f{pile}[2]}~ 
\underbrace{\underset{3}{\back} \, \underset{3}{\back}}_{\f{pile}[3]}~ 
~\cdots~ 
\underbrace{\underset{n}{\back} \, \underset{n}{\back}}_{\f{pile}[n]}~ .
\]
Apply $\PSS_{(n,2)}$ to $(\f{pile}[1], \f{pile}[2] ,\ldots,\f{pile}[n])$ and then we obtain the card-sequence as follows:
\[
\underbrace{\back \, \back \, \back \, \cdots \, \back}_{\f{x}} ~ 
\underset{\alpha_1}{\back} \, \underset{\alpha_1}{\back}
~
\underset{\alpha_2}{\back} \, \underset{\alpha_2}{\back}
~
\underset{\alpha_3}{\back} \, \underset{\alpha_3}{\back}
~
\cdots
~
\underset{\alpha_n}{\back} \, \underset{\alpha_n}{\back}~,
\]
where $\{\alpha_1,\alpha_2,\ldots,\alpha_n\}=\{1,2,\ldots, n\}$.

\item[(3)] Arrange the card-sequence as follows:
\[ 
\underbrace{\back \, \back \, \back \, \cdots \, \back}_{\f{x}} ~
\underset{\alpha_1}{\back} \, \underset{\alpha_2}{\back} \, \underset{\alpha_3}{\back} \, \cdots \, \underset{\alpha_{n-1}}{\back} \, \underset{\alpha_n}{\back} ~~
\underset{\alpha_2}{\back} \, \underset{\alpha_3}{\back} \, \underset{\alpha_4}{\back} \, \cdots \, \underset{\alpha_n}{\back} \, \underset{\alpha_1}{\back}\,.
\]
Apply $\PSS_{(2,n)}$ to the rightmost card-sequence of $2n$ cards as follows:
\[
\bigg|~
\underset{\alpha_1}{\back} \, \underset{\alpha_2}{\back} \, \underset{\alpha_3}{\back} \, \cdots \, \underset{\alpha_{n-1}}{\back} \, \underset{\alpha_n}{\back}
~\bigg|~
\underset{\alpha_2}{\back} \, \underset{\alpha_3}{\back} \, \underset{\alpha_4}{\back} \, \cdots \, \underset{\alpha_n}{\back} \, \underset{\alpha_1}{\back}
~\bigg|.\]
Then we obtain the following card-sequence:
\[\underbrace{\back\, \back\,  \cdots\, \back}_{\f{x}} ~~\underset{\beta_1}{\back} \, \underset{\beta_2}{\back} \, \underset{\beta_3}{\back} \, \cdots \, \underset{\beta_n}{\back}
~~
\underset{\gamma_1}{\back} \, \underset{\gamma_2}{\back} \, \underset{\gamma_3}{\back} \, \cdots \, \underset{\gamma_n}{\back}~,
\]
where $\{(\alpha_1,\alpha_2,\ldots,\alpha_n), (\alpha_2,\ldots,\alpha_n,\alpha_1)\}=\{(\beta_1,\beta_2,\ldots, \beta_n), (\gamma_1,\gamma_2,\ldots,\gamma_n)\}$.

\item[(4)] Arrange the card-sequence as follows:
\[ 
\underset{\f{x}_1}{\back} \,
\underset{\beta_1}{\back} \, \underset{\gamma_1}{\back} ~
\underset{\f{x}_2}{\back} \,
\underset{\beta_2}{\back} \, \underset{\gamma_2}{\back} ~
\underset{\f{x}_3}{\back} \,
\underset{\beta_3}{\back} \, \underset{\gamma_3}{\back} ~
\cdots ~
\underset{\f{x}_n}{\back} \,
\underset{\beta_n}{\back} \, \underset{\gamma_n}{\back}\,.
\]

\item[(5)] Apply $\PSS_{(n,3)}$ to the card-sequence as follows:
\[
\begin{tabular}{|c|c|c|c|c|}
$\back\, \back\, \back$ &
$\back\, \back\, \back$ &
$\back\, \back\, \back$ &
$\cdots$ &
$\back\, \back\, \back$
\end{tabular}\,.
\]

\item[(6)] For all piles, turn over the second and third cards. Let $a_i, b_i \in \{1,2,\ldots, n\}$ be the opened symbols of the second and third cards, respectively, in the $i$-th pile as follows:
\[
\back\,\crd{$a_1$}\,\crd{$b_1$}~~\back\,\crd{$a_2$}\,\crd{$b_2$}~~\back\,\crd{$a_3$}\,\crd{$b_3$}~~\cdots~~\back\,\crd{$a_n$}\,\crd{$b_n$}~\md{.}
\]

\item[(7)] Arrange $n$ piles so that $(c_1, d_1) = (a_1, b_1)$, $d_i = c_{i+1}$, $(1 \leq i \leq n-1)$, and $d_n = c_1$ as follows:
\[
\underset{\f{y}_1}{\back} \,\crd{$c_1$}\,\crd{$d_1$}~~\underset{\f{y}_2}{\back} \,\crd{$c_2$}\,\crd{$d_2$}~~\underset{\f{y}_3}{\back} \,\crd{$c_3$}\,\crd{$d_3$}~~\cdots~~\underset{\f{y}_n}{\back} \,\crd{$c_n$}\,\crd{$d_n$}~.
\]
Then arrange the card-sequence as follows:
\[
\underset{\f{y}_1}{\back} \, \underset{\f{y}_2}{\back} \, \underset{\f{y}_3}{\back} \, \cdots \, \underset{\f{y}_n}{\back} ~ 
\underbrace{\crd{1}\,\crd{1}~\crd{2}\,\crd{2}~\crd{3}\,\crd{3}~\cdots\crd{$n$}\,\crd{$n$}}_{\f{h}}~. 
\]
Then the output card-sequence is $(\f{y}_1,\f{y}_2,\ldots, \f{y}_n)$.
\end{enumerate}

\md{We first show the correctness of the protocol. 
Let $\mathsf{x}=(\mathsf{x}_1,\ldots, \mathsf{x}_n)$ be an input sequence. 
Assume that the protocol outputs the sequence $\mathsf{y}=(\mathsf{y}_1,\ldots, \mathsf{y}_n)$ when $\mathsf{x}$ is given as input. 
Observe that if we apply a graph shuffle for $C_n$ to $\mathsf{x}$, the output sequence is one of the following sequences}
\[ \md{(\mathsf{x}_k,\mathsf{x}_{k+1},\ldots, \mathsf{x}_n, \mathsf{x}_1,\mathsf{x}_2\ldots, \mathsf{x}_{k-1}),\quad  (\mathsf{x}_k,\mathsf{x}_{k-1},\ldots, \mathsf{x}_1, \mathsf{x}_n,\mathsf{x}_{n-1},\ldots, \mathsf{x}_{k+1})}\]
\md{for some $k\in \{1,2,\ldots, n \}$. 
We denote by $\mathsf{Cyc}(k)$ and $\mathsf{Rev}(k)$ the former sequence and the latter sequence, respectively. 
To show the correctness of the protocol, we see that $\mathsf{y}$ is one of $\mathsf{Cyc}(k)$ and $\mathsf{Rev}(k)$ for some $k=1,\ldots, n$.
For $i=1,\ldots, n$, we set $P_i=(\mathsf{x}_i, \beta_i, \gamma_i)$ and put $P=(P_1,\ldots, P_n)$. 
Suppose that $(\alpha_1,\ldots, \alpha_n)$ is equal to $(\beta_1,\ldots, \beta_n)$ in Step (3). 
In this case, it holds $\gamma_i=\beta_{i+1}$ for any $i\in\{1,2,\ldots,n\}$, where $\beta_{n+1}=\beta_1$. 
It follows from the above equations and the argument in the proof of the correctness of the protocol in Subsection \ref{4-1} that $\mathsf{y}=\mathsf{Cyc}(k)$ for some $k$. Similarly, if $(\alpha_1,\ldots, \alpha_n)=(\gamma_1,\ldots, \gamma_n)$ in Step (3), the equations $\gamma_i=\beta_{n-i+1}$ hold for any $i\in\{1,2,\ldots,n\}$. This implies that $\mathsf{y}=\mathsf{Rev}(k)$ for some $k$.}

\md{Next, we show that the distribution of $\mathsf{y}$ is uniform. Assume that $n=2$. We note that $\mathsf{Cyc}(1)=\mathsf{Rev}(1)$ and $\mathsf{Cyc}(2)=\mathsf{Rev}(2)$. Then the candidates appearing as a result of Step (4) are: }
\[ \md{\underset{\f{x}_1}{\back} \,
\underset{1}{\back} \, \underset{2}{\back} ~
\underset{\f{x}_2}{\back} \,
\underset{2}{\back} \, \underset{1}{\back} ~}, \quad 
\md{\underset{\f{x}_1}{\back} \,
\underset{2}{\back} \, \underset{1}{\back} ~
\underset{\f{x}_2}{\back} \,
\underset{1}{\back} \, \underset{2}{\back} ~}\ ,\] 
\md{and these each have a probability of $\dfrac{1}{2}$. Thus, the probabilities that $\mathsf{y}=(\mathsf{x}_1,\mathsf{x}_2)$ and $\mathsf{y}=(\mathsf{x}_2,\mathsf{x}_1)$ are same. Now, we assume that $n\geq 3$. In this case, for any $k=1,\ldots, n$, all sequences $\mathsf{Cyc}(k)$ and $\mathsf{Rev}(k)$ are distinct. In order to get $\mathsf{y}=\mathsf{Cyc}(k)$,  it requires that $(\alpha_1,\ldots, \alpha_n)=(\beta_1,\ldots, \beta_n)$ in Step (3) and $\sigma^{-1}(1)=k$, where $\sigma$ is the chosen permutation in Step (5). 
Hence, the probability that $\mathsf{y}=\mathsf{Cyc}(k)$ is $\dfrac{1}{2k}$. Similarly, the probability that $\mathsf{y}=\mathsf{Rev}(k)$ is also $\dfrac{1}{2k}$. This shows that the protocol is correct.}

\md{We show the correctness of the protocol. 
Assume that $\sigma\in\mathfrak{S}_n$ and $\tau\in \mathfrak{S}_n$ are chosen in Step (2) and Step (5), respectively. Then the first card in the $i$-th pile in Step (6) is $\mathsf{x}_{\tau^{-1}(i)}$. On the other hand, the second and third cards in the $i$-th pile in Step (5) are depending on the result of Step (3), and they are determined as follows. 
If $(\alpha_1,\ldots, \alpha_n)=(\beta_1,\ldots, \beta_n)$, then $a_i=\tau^{-1}\sigma^{-1}(i)$ and $b_i=\tau^{-1}\sigma^{-1}(i+1)$, 
otherwise, $a_i=\tau^{-1}\sigma^{-1}(i+1)$ and $b_i=\tau^{-1}\sigma^{-1}(i)$. Here, we consider $n+1$ as $1$. 
In either case,  these open symbols $a_1,\ldots, a_n$ and $b_1,\ldots, b_n$ do not allow us to guess the first card of any pile since $\sigma$ is chosen uniformly at random in Step (5). Therefore, the protocol is secure.}

\section{Conclusions and Future Works}

\md{In this paper, we show that any graph shuffle can be done by PSSs. 
In particular, we need $2(n+m)$ cards and $|\f{Deg}_G|+1$ PSSs, where $n$ and $m$ are the numbers of vertices and arrows of $G$, respectively. 
We left as open problems (1) to remove the computation of an isomorphism between two isomorphic graphs in a graph shuffle protocol keeping everything efficient and (2) to find another interesting applications for our graph shuffle protocol. 
We hope that this research direction (i.e., constructing a nontrivial shuffle from the standard shuffles such as RCs, RBCs, and PSSs) will attract the interest of researchers on card-based cryptography and new shuffle protocols will be proposed in future work. 
}

\bibliographystyle{abbrv}
\bibliography{card}

\begin{thebibliography}{10}

\bibitem{AbeAPKC18}
Y.~Abe, Y.~Hayashi, T.~Mizuki, and H.~Sone.
\newblock Five-card {AND} protocol in committed format using only practical
  shuffles.
\newblock In K.~Emura, J.~H. Seo, and Y.~Watanabe, editors, {\em Proceedings of
  the 5th {ACM} on {ASIA} Public-Key Cryptography Workshop, APKC@AsiaCCS,
  Incheon, Republic of Korea, June 4, 2018}, pages 3--8. {ACM}, 2018.

\bibitem{AbeNGC21}
Y.~Abe, Y.~Hayashi, T.~Mizuki, and H.~Sone.
\newblock Five-card {AND} computations in committed format using only uniform
  cyclic shuffles.
\newblock {\em New Gener. Comput.}, 39(1):97--114, 2021.

\bibitem{PascalFUN16}
X.~Bultel, J.~Dreier, J.~Dumas, and P.~Lafourcade.
\newblock Physical zero-knowledge proofs for akari, takuzu, kakuro and kenken.
\newblock In E.~D. Demaine and F.~Grandoni, editors, {\em 8th International
  Conference on Fun with Algorithms, {FUN} 2016, June 8-10, 2016, La Maddalena,
  Italy}, volume~49 of {\em LIPIcs}, pages 8:1--8:20. Schloss Dagstuhl -
  Leibniz-Zentrum f{\"{u}}r Informatik, 2016.

\bibitem{ShinagawaSSS18}
X.~Bultel, J.~Dreier, J.~Dumas, P.~Lafourcade, D.~Miyahara, T.~Mizuki,
  A.~Nagao, T.~Sasaki, K.~Shinagawa, and H.~Sone.
\newblock Physical zero-knowledge proof for makaro.
\newblock In T.~Izumi and P.~Kuznetsov, editors, {\em Stabilization, Safety,
  and Security of Distributed Systems - 20th International Symposium, {SSS}
  2018, Tokyo, Japan, November 4-7, 2018, Proceedings}, volume 11201 of {\em
  Lecture Notes in Computer Science}, pages 111--125. Springer, 2018.

\bibitem{CHL13}
E.~Cheung, C.~Hawthorne, and P.~Lee.
\newblock Cs 758 project: Secure computation with playing cards, 2013.
\newblock \url{https://cdchawthorne.com/writings/secure_playing_cards.pdf}.

\bibitem{Kilian93}
C.~Cr{\'{e}}peau and J.~Kilian.
\newblock Discreet solitary games.
\newblock In D.~R. Stinson, editor, {\em Advances in Cryptology - {CRYPTO} '93,
  13th Annual International Cryptology Conference, Santa Barbara, California,
  USA, August 22-26, 1993, Proceedings}, volume 773 of {\em Lecture Notes in
  Computer Science}, pages 319--330. Springer, 1993.

\bibitem{Boer89}
B.~den Boer.
\newblock More efficient match-making and satisfiability: \emph{The Five Card
  Trick}.
\newblock In J.~Quisquater and J.~Vandewalle, editors, {\em Advances in
  Cryptology - {EUROCRYPT} '89, Workshop on the Theory and Application of of
  Cryptographic Techniques, Houthalen, Belgium, April 10-13, 1989,
  Proceedings}, volume 434 of {\em Lecture Notes in Computer Science}, pages
  208--217. Springer, 1989.

\bibitem{DumasCOCOON19}
J.~Dumas, P.~Lafourcade, D.~Miyahara, T.~Mizuki, T.~Sasaki, and H.~Sone.
\newblock Interactive physical zero-knowledge proof for norinori.
\newblock In D.~Du, Z.~Duan, and C.~Tian, editors, {\em Computing and
  Combinatorics - 25th International Conference, {COCOON} 2019, Xi'an, China,
  July 29-31, 2019, Proceedings}, volume 11653 of {\em Lecture Notes in
  Computer Science}, pages 166--177. Springer, 2019.

\bibitem{CLZ}
L.~L. G.~Chartrand and P.~Zhang.
\newblock {\em Graphs \& Digraphs (six edition)}.
\newblock CRC Press, 2015.

\bibitem{GradwohlFUN07}
R.~Gradwohl, M.~Naor, B.~Pinkas, and G.~N. Rothblum.
\newblock Cryptographic and physical zero-knowledge proof systems for solutions
  of sudoku puzzles.
\newblock In {\em Fun with Algorithms, 4th International Conference, {FUN}
  2007, Castiglioncello, Italy, June 3-5, 2007, Proceedings}, pages 166--182,
  2007.

\bibitem{HashimotoICITS17}
Y.~Hashimoto, K.~Shinagawa, K.~Nuida, M.~Inamura, and G.~Hanaoka.
\newblock Secure grouping protocol using a deck of cards.
\newblock In J.~Shikata, editor, {\em Information Theoretic Security - 10th
  International Conference, {ICITS} 2017, Hong Kong, China, November 29 -
  December 2, 2017, Proceedings}, volume 10681 of {\em Lecture Notes in
  Computer Science}, pages 135--152. Springer, 2017.

\bibitem{Heather14}
J.~Heather, S.~Schneider, and V.~Teague.
\newblock Cryptographic protocols with everyday objects.
\newblock {\em Formal Asp. Comput.}, 26(1):37--62, 2014.

\bibitem{IshikawaUCNC15}
R.~Ishikawa, E.~Chida, and T.~Mizuki.
\newblock Efficient card-based protocols for generating a hidden random
  permutation without fixed points.
\newblock In C.~S. Calude and M.~J. Dinneen, editors, {\em Unconventional
  Computation and Natural Computation - 14th International Conference, {UCNC}
  2015, Auckland, New Zealand, August 30 - September 3, 2015, Proceedings},
  volume 9252 of {\em Lecture Notes in Computer Science}, pages 215--226.
  Springer, 2015.

\bibitem{Koch18b}
A.~Koch and S.~Walzer.
\newblock Private function evaluation with cards.
\newblock {\em {IACR} Cryptology ePrint Archive}, 2018:1113, 2018.

\bibitem{KochFUN21}
A.~Koch and S.~Walzer.
\newblock Foundations for actively secure card-based cryptography.
\newblock In M.~Farach{-}Colton, G.~Prencipe, and R.~Uehara, editors, {\em 10th
  International Conference on Fun with Algorithms, {FUN} 2021, May 30 to June
  1, 2021, Favignana Island, Sicily, Italy}, volume 157 of {\em LIPIcs}, pages
  17:1--17:23. Schloss Dagstuhl - Leibniz-Zentrum f{\"{u}}r Informatik, 2021.

\bibitem{KoyamaCSR21}
H.~Koyama, D.~Miyahara, T.~Mizuki, and H.~Sone.
\newblock A secure three-input {AND} protocol with a standard deck of minimal
  cards.
\newblock In R.~Santhanam and D.~Musatov, editors, {\em Computer Science -
  Theory and Applications - 16th International Computer Science Symposium in
  Russia, {CSR} 2021, Sochi, Russia, June 28 - July 2, 2021, Proceedings},
  volume 12730 of {\em Lecture Notes in Computer Science}, pages 242--256.
  Springer, 2021.

\bibitem{KoyamaAPKC21}
H.~Koyama, K.~Toyoda, D.~Miyahara, and T.~Mizuki.
\newblock New card-based copy protocols using only random cuts.
\newblock In K.~Emura and Y.~Wang, editors, {\em Proceedings of the 8th on
  {ASIA} Public-Key Cryptography Workshop, APKC@AsiaCCS 2021, Virtual Event
  Hong Kong, 7 June, 2021}, pages 13--22. {ACM}, 2021.

\bibitem{MiyaharaISPEC19}
P.~Lafourcade, D.~Miyahara, T.~Mizuki, T.~Sasaki, and H.~Sone.
\newblock A physical {ZKP} for slitherlink: How to perform physical
  topology-preserving computation.
\newblock In S.~Heng and J.~L{\'{o}}pez, editors, {\em Information Security
  Practice and Experience - 15th International Conference, {ISPEC} 2019, Kuala
  Lumpur, Malaysia, November 26-28, 2019, Proceedings}, volume 11879 of {\em
  Lecture Notes in Computer Science}, pages 135--151. Springer, 2019.

\bibitem{Lindell20}
Y.~Lindell.
\newblock Secure multiparty computation (mpc).
\newblock Cryptology ePrint Archive, Report 2020/300, 2020.
\newblock \url{https://ia.cr/2020/300}.

\bibitem{Marcedone15}
A.~Marcedone, Z.~Wen, and E.~Shi.
\newblock Secure dating with four or fewer cards.
\newblock Cryptology ePrint Archive, Report 2015/1031, 2015.

\bibitem{Nauty}
B.~McKay and A.~Piperno.
\newblock The nauty traces page.

\bibitem{MiyaharaCOCOA18}
D.~Miyahara, Y.~Hayashi, T.~Mizuki, and H.~Sone.
\newblock Practical and easy-to-understand card-based implementation of yao's
  millionaire protocol.
\newblock In D.~Kim, R.~N. Uma, and A.~Zelikovsky, editors, {\em Combinatorial
  Optimization and Applications - 12th International Conference, {COCOA} 2018,
  Atlanta, GA, USA, December 15-17, 2018, Proceedings}, volume 11346 of {\em
  Lecture Notes in Computer Science}, pages 246--261. Springer, 2018.

\bibitem{MiyaharaTCS20}
D.~Miyahara, Y.~Hayashi, T.~Mizuki, and H.~Sone.
\newblock Practical card-based implementations of yao's millionaire protocol.
\newblock {\em Theor. Comput. Sci.}, 803:207--221, 2020.

\bibitem{MiyaharaFUN21}
D.~Miyahara, L.~Robert, P.~Lafourcade, S.~Takeshige, T.~Mizuki, K.~Shinagawa,
  A.~Nagao, and H.~Sone.
\newblock Card-based {ZKP} protocols for takuzu and juosan.
\newblock In M.~Farach{-}Colton, G.~Prencipe, and R.~Uehara, editors, {\em 10th
  International Conference on Fun with Algorithms, {FUN} 2021, May 30 to June
  1, 2021, Favignana Island, Sicily, Italy}, volume 157 of {\em LIPIcs}, pages
  20:1--20:21. Schloss Dagstuhl - Leibniz-Zentrum f{\"{u}}r Informatik, 2021.

\bibitem{MizukiTeaching16}
T.~Mizuki.
\newblock Applications of card-based cryptography to education.
\newblock {\em IEICE Technical Report}, 116(289):13--17, 2016.
\newblock {(In Japanese)}.

\bibitem{MizukiTCS16}
T.~Mizuki.
\newblock Card-based protocols for securely computing the conjunction of
  multiple variables.
\newblock {\em Theor. Comput. Sci.}, 622:34--44, 2016.

\bibitem{MizukiCANS16}
T.~Mizuki.
\newblock Efficient and secure multiparty computations using a standard deck of
  playing cards.
\newblock In S.~Foresti and G.~Persiano, editors, {\em Cryptology and Network
  Security - 15th International Conference, {CANS} 2016, Milan, Italy, November
  14-16, 2016, Proceedings}, volume 10052 of {\em Lecture Notes in Computer
  Science}, pages 484--499, 2016.

\bibitem{MizukiUCNC13}
T.~Mizuki, I.~K. Asiedu, and H.~Sone.
\newblock Voting with a logarithmic number of cards.
\newblock In G.~Mauri, A.~Dennunzio, L.~Manzoni, and A.~E. Porreca, editors,
  {\em Unconventional Computation and Natural Computation - 12th International
  Conference, {UCNC} 2013, Milan, Italy, July 1-5, 2013. Proceedings}, volume
  7956 of {\em Lecture Notes in Computer Science}, pages 162--173. Springer,
  2013.

\bibitem{MizukiAC12}
T.~Mizuki, M.~Kumamoto, and H.~Sone.
\newblock The five-card trick can be done with four cards.
\newblock In X.~Wang and K.~Sako, editors, {\em Advances in Cryptology -
  {ASIACRYPT} 2012 - 18th International Conference on the Theory and
  Application of Cryptology and Information Security, Beijing, China, December
  2-6, 2012. Proceedings}, volume 7658 of {\em Lecture Notes in Computer
  Science}, pages 598--606. Springer, 2012.

\bibitem{MizukiIJISEC14}
T.~Mizuki and H.~Shizuya.
\newblock A formalization of card-based cryptographic protocols via abstract
  machine.
\newblock {\em Int. J. Inf. Sec.}, 13(1):15--23, 2014.

\bibitem{MizukiFAW09}
T.~Mizuki and H.~Sone.
\newblock Six-card secure {AND} and four-card secure {XOR}.
\newblock In X.~Deng, J.~E. Hopcroft, and J.~Xue, editors, {\em Frontiers in
  Algorithmics, Third International Workshop, {FAW} 2009, Hefei, China, June
  20-23, 2009. Proceedings}, volume 5598 of {\em Lecture Notes in Computer
  Science}, pages 358--369. Springer, 2009.

\bibitem{Uchiike06}
T.~Mizuki, F.~Uchiike, and H.~Sone.
\newblock Securely computing {XOR} with 10 cards.
\newblock {\em The Australasian Journal of Combinatorics}, 36:279--293, 2006.

\bibitem{MurataWALCOM21}
S.~Murata, D.~Miyahara, T.~Mizuki, and H.~Sone.
\newblock Efficient generation of a card-based uniformly distributed random
  derangement.
\newblock In R.~Uehara, S.~Hong, and S.~C. Nandy, editors, {\em {WALCOM:}
  Algorithms and Computation - 15th International Conference and Workshops,
  {WALCOM} 2021, Yangon, Myanmar, February 28 - March 2, 2021, Proceedings},
  volume 12635 of {\em Lecture Notes in Computer Science}, pages 78--89.
  Springer, 2021.

\bibitem{Niemi98}
V.~Niemi and A.~Renvall.
\newblock Secure multiparty computations without computers.
\newblock {\em Theor. Comput. Sci.}, 191(1-2):173--183, 1998.

\bibitem{Niemi99}
V.~Niemi and A.~Renvall.
\newblock Solitaire zero-knowledge.
\newblock {\em Fundam. Informaticae}, 38(1-2):181--188, 1999.

\bibitem{NishidaTAMC15}
T.~Nishida, Y.~Hayashi, T.~Mizuki, and H.~Sone.
\newblock Card-based protocols for any boolean function.
\newblock In R.~Jain, S.~Jain, and F.~Stephan, editors, {\em Theory and
  Applications of Models of Computation - 12th Annual Conference, {TAMC} 2015,
  Singapore, May 18-20, 2015, Proceedings}, volume 9076 of {\em Lecture Notes
  in Computer Science}, pages 110--121. Springer, 2015.

\bibitem{NishidaIEICE15}
T.~Nishida, Y.~Hayashi, T.~Mizuki, and H.~Sone.
\newblock Securely computing three-input functions with eight cards.
\newblock {\em {IEICE} Transactions}, 98-A(6):1145--1152, 2015.

\bibitem{NishidaTPNC13}
T.~Nishida, T.~Mizuki, and H.~Sone.
\newblock Securely computing the three-input majority function with eight
  cards.
\newblock In A.~Dediu, C.~Mart{\'{\i}}n{-}Vide, B.~Truthe, and M.~A.
  Vega{-}Rodr{\'{\i}}guez, editors, {\em Theory and Practice of Natural
  Computing - Second International Conference, {TPNC} 2013, C{\'{a}}ceres,
  Spain, December 3-5, 2013, Proceedings}, volume 8273 of {\em Lecture Notes in
  Computer Science}, pages 193--204. Springer, 2013.

\bibitem{RobertSSS20}
L.~Robert, D.~Miyahara, P.~Lafourcade, and T.~Mizuki.
\newblock Physical zero-knowledge proof for suguru puzzle.
\newblock In S.~Devismes and N.~Mittal, editors, {\em Stabilization, Safety,
  and Security of Distributed Systems - 22nd International Symposium, {SSS}
  2020, Austin, TX, USA, November 18-21, 2020, Proceedings}, volume 12514 of
  {\em Lecture Notes in Computer Science}, pages 235--247. Springer, 2020.

\bibitem{RobertCiE21}
L.~Robert, D.~Miyahara, P.~Lafourcade, and T.~Mizuki.
\newblock Interactive physical {ZKP} for connectivity: Applications to nurikabe
  and hitori.
\newblock In L.~D. Mol, A.~Weiermann, F.~Manea, and D.~Fern{\'{a}}ndez{-}Duque,
  editors, {\em Connecting with Computability - 17th Conference on
  Computability in Europe, CiE 2021, Virtual Event, Ghent, July 5-9, 2021,
  Proceedings}, volume 12813 of {\em Lecture Notes in Computer Science}, pages
  373--384. Springer, 2021.

\bibitem{RuangwisesNGC21}
S.~Ruangwises and T.~Itoh.
\newblock Physical zero-knowledge proof for numberlink puzzle and k
  vertex-disjoint paths problem.
\newblock {\em New Gener. Comput.}, 39(1):3--17, 2021.

\bibitem{RuangwisesTCS21a}
S.~Ruangwises and T.~Itoh.
\newblock Physical zero-knowledge proof for ripple effect.
\newblock {\em Theor. Comput. Sci.}, 895:115--123, 2021.

\bibitem{RuangwisesTCS21}
S.~Ruangwises and T.~Itoh.
\newblock Securely computing the \emph{n}-variable equality function with
  2\emph{n} cards.
\newblock {\em Theor. Comput. Sci.}, 887:99--110, 2021.

\bibitem{SaitoTPNC20}
T.~Saito, D.~Miyahara, Y.~Abe, T.~Mizuki, and H.~Shizuya.
\newblock How to implement a non-uniform or non-closed shuffle.
\newblock In C.~Mart{\'{\i}}n{-}Vide, M.~A. Vega{-}Rodr{\'{\i}}guez, and
  M.~Yang, editors, {\em Theory and Practice of Natural Computing - 9th
  International Conference, {TPNC} 2020, Taoyuan, Taiwan, December 7-9, 2020,
  Proceedings}, volume 12494 of {\em Lecture Notes in Computer Science}, pages
  107--118. Springer, 2020.

\bibitem{SasakiTCS20}
T.~Sasaki, D.~Miyahara, T.~Mizuki, and H.~Sone.
\newblock Efficient card-based zero-knowledge proof for sudoku.
\newblock {\em Theor. Comput. Sci.}, 839:135--142, 2020.

\bibitem{SasakiFUN18}
T.~Sasaki, T.~Mizuki, and H.~Sone.
\newblock Card-based zero-knowledge proof for sudoku.
\newblock In H.~Ito, S.~Leonardi, L.~Pagli, and G.~Prencipe, editors, {\em 9th
  International Conference on Fun with Algorithms, {FUN} 2018, June 13-15,
  2018, La Maddalena, Italy}, volume 100 of {\em LIPIcs}, pages 29:1--29:10.
  Schloss Dagstuhl - Leibniz-Zentrum f{\"{u}}r Informatik, 2018.

\bibitem{ShinagawaICISC18}
K.~Shinagawa and T.~Mizuki.
\newblock The six-card trick: Secure computation of three-input equality.
\newblock In K.~Lee, editor, {\em Information Security and Cryptology - {ICISC}
  2018 - 21st International Conference, Seoul, South Korea, November 28-30,
  2018, Revised Selected Papers}, volume 11396 of {\em Lecture Notes in
  Computer Science}, pages 123--131. Springer, 2018.

\bibitem{ShinagawaFAW19}
K.~Shinagawa and T.~Mizuki.
\newblock Secure computation of any boolean function based on any deck of
  cards.
\newblock In Y.~Chen, X.~Deng, and M.~Lu, editors, {\em Frontiers in
  Algorithmics - 13th International Workshop, {FAW} 2019, Sanya, China, April
  29 - May 3, 2019, Proceedings}, volume 11458 of {\em Lecture Notes in
  Computer Science}, pages 63--75. Springer, 2019.

\bibitem{ShinagawaDAM21}
K.~Shinagawa and K.~Nuida.
\newblock A single shuffle is enough for secure card-based computation of any
  boolean circuit.
\newblock {\em Discret. Appl. Math.}, 289:248--261, 2021.

\bibitem{ShinodaSecITC20}
Y.~Shinoda, D.~Miyahara, K.~Shinagawa, T.~Mizuki, and H.~Sone.
\newblock Card-based covert lottery.
\newblock In D.~Maimut, A.~Oprina, and D.~Sauveron, editors, {\em Innovative
  Security Solutions for Information Technology and Communications - 13th
  International Conference, SecITC 2020, Bucharest, Romania, November 19-20,
  2020, Revised Selected Papers}, volume 12596 of {\em Lecture Notes in
  Computer Science}, pages 257--270. Springer, 2020.

\bibitem{Stiglic01}
A.~Stiglic.
\newblock Computations with a deck of cards.
\newblock {\em Theor. Comput. Sci.}, 259(1-2):671--678, 2001.

\bibitem{TakashimaCOCOA19}
K.~Takashima, Y.~Abe, T.~Sasaki, D.~Miyahara, K.~Shinagawa, T.~Mizuki, and
  H.~Sone.
\newblock Card-based secure ranking computations.
\newblock In Y.~Li, M.~Cardei, and Y.~Huang, editors, {\em Combinatorial
  Optimization and Applications - 13th International Conference, {COCOA} 2019,
  Xiamen, China, December 13-15, 2019, Proceedings}, volume 11949 of {\em
  Lecture Notes in Computer Science}, pages 461--472. Springer, 2019.

\bibitem{TakashimaTCS20}
K.~Takashima, Y.~Abe, T.~Sasaki, D.~Miyahara, K.~Shinagawa, T.~Mizuki, and
  H.~Sone.
\newblock Card-based protocols for secure ranking computations.
\newblock {\em Theor. Comput. Sci.}, 845:122--135, 2020.

\bibitem{TakashimaTPNC19}
K.~Takashima, D.~Miyahara, T.~Mizuki, and H.~Sone.
\newblock Card-based protocol against actively revealing card attack.
\newblock In C.~Mart{\'{\i}}n{-}Vide, G.~T. Pond, and M.~A.
  Vega{-}Rodr{\'{\i}}guez, editors, {\em Theory and Practice of Natural
  Computing - 8th International Conference, {TPNC} 2019, Kingston, ON, Canada,
  December 9-11, 2019, Proceedings}, volume 11934 of {\em Lecture Notes in
  Computer Science}, pages 95--106. Springer, 2019.

\bibitem{ToyodaAPKC20}
K.~Toyoda, D.~Miyahara, T.~Mizuki, and H.~Sone.
\newblock Six-card finite-runtime {XOR} protocol with only random cut.
\newblock In K.~Emura and N.~Yanai, editors, {\em Proceedings of the 7th on
  {ASIA} Public-Key Cryptography Workshop, APKC@AsiaCCS 2020, Taipei, Taiwan,
  October 6, 2020}, pages 2--8. {ACM}, 2020.

\bibitem{Yao82}
A.~C. Yao.
\newblock Protocols for secure computations (extended abstract).
\newblock In {\em 23rd Annual Symposium on Foundations of Computer Science,
  Chicago, Illinois, USA, 3-5 November 1982}, pages 160--164. {IEEE} Computer
  Society, 1982.

\bibitem{Yao86}
A.~C. Yao.
\newblock How to generate and exchange secrets (extended abstract).
\newblock In {\em 27th Annual Symposium on Foundations of Computer Science,
  Toronto, Canada, 27-29 October 1986}, pages 162--167. {IEEE} Computer
  Society, 1986.

\end{thebibliography}

\section*{Declarations}

\subsection*{Funding}
K. Miyamoto was partly supported by JSPS KAKENHI 20K14302. 
K. Shinagawa was partly supported by JSPS KAKENHI 21K17702. 

\end{document}